\documentclass[aip,pof,amsmath,amssymb]{revtex4-1}
\usepackage{hyperref}
\hypersetup{
    colorlinks,
    linktocpage,
    citecolor=blue,
    filecolor=black,
    linkcolor=blue,
    urlcolor=blue,
}
\usepackage{graphicx,subfig} 
\usepackage{float}
\usepackage{graphicx}
\usepackage{tikz}
\usepackage{pagecolor}
\usepackage{ulem}
\usepackage{array}
\usepackage{multirow}
\usepackage{calc}       
\usepackage{multirow}
\usepackage{colortbl}
\draft 

\begin{document}
\title{A hybrid discrete exterior calculus and finite difference method for anelastic convection in spherical shells } 
\author{Hamid Hassan Khan} \affiliation{Mechanical Engineering, Physical Science and Engineering Division (PSE), King Abdullah University of Science and Technology, Thuwal, 23955-6900, Saudi Arabia} 
\author{Pankaj Jagad} \affiliation{Mechanical Engineering, Physical Science and Engineering Division (PSE), King Abdullah University of Science and Technology, Thuwal, 23955-6900, Saudi Arabia}
\author{Matteo Parsani} \affiliation{Mechanical Engineering, Physical Science and Engineering Division (PSE), King Abdullah University of Science and Technology, Thuwal, 23955-6900, Saudi Arabia}

\begin{abstract}
\textbf{Abstract}\\
The present work develops, verifies, and benchmarks a hybrid discrete exterior calculus and finite difference (DEC-FD) method for density-stratified thermal convection in spherical shells. Discrete exterior calculus (DEC) is notable for its coordinate independence and structure preservation properties. The hybrid DEC-FD method for Boussinesq convection has been developed by Mantravadi et al. \cite{mantravadi2023hybrid}. Motivated by astrophysics problems, we extend this method assuming anelastic convection, which retains density stratification; this has been widely used for decades to understand thermal convection in stars and giant planets. In the present work, the governing equations are splitted into surface and radial components and discrete anelastic equations are derived by replacing spherical surface operators with DEC and radial operators with FD operators. The novel feature of this work is the discretization of anelastic equations with the DEC-FD method and the assessment of a hybrid solver for density-stratified thermal convection in spherical shells. The discretized anelastic equations are verified using the method of manufactured solution (MMS). We performed a series of three-dimensional convection simulations in a spherical shell geometry and examined the effect of density ratio on convective flow structures and energy dynamics. The present observations are in agreement with the benchmark models.
\\

\textbf{Keywords}: Anelastic approximation, density-stratification, discrete exterior calculus (DEC), finite difference method, spherical shells, solar convection
\end{abstract}
\maketitle 
\section{Introduction}
Thermal convection is a physical phenomenon found in astrophysical objects. It is a significant transport mechanism that transfers energy across planetary and stellar interiors. In giant stars, the transport phenomena occur in the convective core, while in the Sun, the transport occurs in the outer convective envelope\cite{schumacher2020colloquium}. Similarly, planets' convective motions are large-scale and feature plate tectonics on Earth \cite{trompert1998mantle,tackley2000mantle,bercovici2003generation}, dynamo-generated magnetic fields, and zonal winds on Jupiter and other planets\cite{heimpel2005simulation,verhoeven2014compressional}.

The giant planets and stars exhibit non-negligible density stratification in their convective regions. In contrast, the heat flux in convective regions is sufficiently small, and the flows are subsonic, resulting in a slight change in density from equilibrium. Due to density stratification in giant planets and stars, the Boussinesq approximation has to be improved for studying thermal convection. On the other hand, using a fully compressible model will increase the computational burden and reduce accuracy due to the presence of sound waves. The anelastic approximation has been widely used for decades to understand thermal convection in stars and giant planets. Anelastic thermal convection, a paradigm for solar convection, retains the effect of density stratification and filters out sound waves from fully compressible flow\cite{jones2011anelastic,verhoeven2015anelastic,kessar2019scale}, which is computationally less expensive.

Ogura and Phillips \cite{ogura1962scale} first derived the anelastic equations in a stable stratified atmosphere. Later, Gough \cite{gough1969anelastic}, Gilman and Glatzmaier \cite{gilman1981compressible}, and Lantz and Fan \cite{lantz1999anelastic} extended the anelastic approximation to convecting atmospheres. Further simplification and a more complete derivation was made independently by Lantz \cite{lantz1992dynamical} and Braginsky and Roberts \cite{braginsky1995equations}. Since then, anelastic approximations have been extensively utilized to model density-stratified systems in planetary and stellar interiors or atmospheres. To study anelastic thermal convection inside giant planets and stars, previous studies \cite{verhoeven2015anelastic,kessar2019scale,jones2022fully,curbelo2019numerical} considered fluids in a planar layer between two parallel plates. Verhoeven et al. \cite{verhoeven2015anelastic} and Curbelo et al. \cite{curbelo2019numerical} reported the performance of anelastic approximation as compared to the compressible model. Curbelo et al. \cite{curbelo2019numerical} illustrated that the structures at top and bottom layers differ at higher Prandtl number.

The numerical study of solar convection is natural and reliable in spherical shell geometry. A more accurate numerical method is required to solve the anelastic convection model compared to the Boussinesq convection model, as the density variation leads to small spatial scales.  Recently, development of anelastic code with the advancement in supercomputing is practical and significantly important to study underlying dynamics. Indeed, the anelastic convection studies in spherical shells are less numerous as compared to planar geometry. However, few anelastic codes for thermal convection in spherical shells are available. Clune et al. \cite{clune1999computational} applied anelastic approximation on governing equations to develop anelastic harmonic code (ASH) for stellar convection in spherical shells. ASH \cite{clune1999computational,miesch2000three} code utilizes spherical harmonic functions in the latitudinal and azimuthal directions, and Chebyshev polynomials or finite differences in the radial direction. MagIC \cite{wicht2002inner, gastine2015turbulent} code employed numerical methods similar to ASH code, while both codes were derived independently from the original version of the Glatzmaier \cite{glatzmaier1984numerical} code. Jones et al. \cite{jones2011anelastic} benchmarked another code slightly different from MagIC and ASH code. Indeed, these four codes - ASH, MagIC, Glatzmaier, and Leeds codes have a common feature of spherical harmonic expansion in the latitudinal and azimuthal direction and Chebyshev polynomials or finite difference in the radial direction. Spherical harmonic expansions are employed using the pseudo-spectra method. In this modern era of parallel computing, the spherical spectra method has advantages and disadvantages. The spherical pseudo-spectra method is accurate and easy to code. However, the spherical harmonic expansion utilizes global integration, leading to global communication and inefficient parallel computing performance. The most challenging problem of the spherical spectra method is that it cannot be easily employed for non-spherical coordinates, the effect of which plays an essential role in dynamo action driven by planetary precession, libration, or tides.  Zhan et al. \cite{zhan2012anelastic} developed a finite element method code for three-dimensional anelastic convection in spherical geometry. More details about the computational approaches for the simulation of stellar convection are reported in review work \cite{kupka2017modelling,schumacher2020colloquium}. 

Simulation and modeling communities have recently begun to embrace discrete exterior calculus (DEC) as a powerful numerical method that preserves the structures and has the benefit of coordinate independent system. The exterior calculus is the calculus of differential form, and the extension of exterior calculus on discrete space is defined as the Discrete exterior calculus (DEC). By incorporating differential forms, DEC-based methods satisfy the physical and mathematical properties of governing equations (partial differential equations). Due to differential forms, the DEC numerical method satisfies orthogonality relations; $\nabla \cdot \nabla \times \textbf{v} = 0$ for any vector \textbf{v} and $\nabla \times \nabla s =0$ for any scalar s. Early application of DEC was limited to computer vision and computer graphics \cite{desbrun2006discrete,desbrun2003discrete,grinspun2006discrete}. Recently, scientists and engineers are incorporating a DEC-based approach in the area of fluid dynamics \cite{hirani2015numerical,mohamed2016discrete,jagad2021primitive,jagad2020investigation,jagad2021effects,wang2022discrete,mohamed2018numerical,schulz2020convergence} and electromagnetism \cite{chen2016discrete}. Moreover, the DEC method can be easily modified for non-spherical geometries such as triaxial ellipsoids and spheroids.

The DEC method is notable for its coordinate independence and structure preservation properties \cite{mohamed2016discrete,jagad2021primitive}. Based on these unique features, Mantravadi et al.\cite{mantravadi2023hybrid} recently developed a hybrid discrete exterior calculus and finite difference (DEC-FD) method for Boussinesq thermal convection in spherical shells. In contrast, in astrophysics problems, the density stratification is non-negligible in the convective regions of the giant planets and stars. Therefore, we extend the hybrid DEC-FD method to density-stratified thermal convection in spherical shells. In the present work, we develop a hybrid DEC-FD solver using an anelastic approximation. The novel feature of the present work is to derive operator-split anelastic equations in spherical/surface and radial operators. Subsequently, the spherical/surface operators are replaced with DEC operators, and radial operators are replaced with FD operators. More specifically, the objectives of the present work are two-fold. First, the formulation and discretization of the anelastic equations using the hybrid DEC-FD method. Second, to verify the formulation and benchmark the hybrid DEC-FD solver for density-stratified thermal convection in spherical shells. The discretized anelastic equations are verified using the method of manufactured solution (MMS). The thermal convection simulations in spherical shells demonstrate the hybrid DEC-FD solver assessment for anelastic approximation. In addition, we examined the effect of density ratio on the convective flow structures (granular patterns) and energy dynamics.

This paper is organized as follows: The governing equations are introduced in section \ref{sec:governing_eq}, where section \ref{sec:anelastic_eq} presents non-dimensional anelastic equations, and section \ref{sec:split_eq} demonstrates the operator split into surface and radial operators. In section \ref{sec:expression}, we derive the anelastic equations in exterior calculus notation. The discretization is demonstrated in section \ref{sec:discretization}, consisting of the domain discretization (section \ref{sec:domain_discretization}), and discretization of anelastic equations in hybrid discrete exterior calculus and finite difference (DEC-FD) framework (section \ref{sec:discretization_anelastic}). The implementation of hybrid DEC-FD solver is presented in section \ref{sec:implementation}. The numerical investigation illustrated in section \ref{sec:numerical_investigation} includes verification (section \ref{sec:MMS}) and validation (section \ref{sec:validation}) of hybrid solver, and analysis of convective flows (section \ref{sec:structures}-\ref{sec:density}). Finally, the conclusions are drawn in section \ref{sec:conclusions}. The discretization details of the viscous heating term are reported in appendix \hyperref[appendixA]{A}. The residue (or source) terms introduced in the governing equations for MMS are reported in appendix \hyperref[appendixB]{B}.

\section{Governing Equations} \label{sec:governing_eq}
\subsection{Anelastic equation} \label{sec:anelastic_eq}
The convection in the sun is modeled with the anelastic approximation.  For anelastic approximation \cite{verhoeven2015anelastic,kessar2019scale,gilman1981compressible,lantz1999anelastic,mizerski2017rigorous}, the thermodynamic variables (density, pressure, and temperature) present in traditional compressible flow equation are decomposed in adiabatic (r-dependent reference state) and superadiabatic (perturbed state) components. The anelastic approximation assumes a nearly adiabatic basic state, which allows entropy-based equations to be formulated; for details on entropy-based formulation, see previous work\cite{gilman1981compressible,lantz1999anelastic,mizerski2017rigorous}. We consider convection, under anelastic approximation, of the internally heated fluid element confined in the spherical shell of inner radius $r_i$ and outer radius $r_o$. Kessar et al. \cite{kessar2019scale} and Mizerski $\&$ Tobias \cite{mizerski2011effect} used a dimensionless set of anelastic equations to study solar convection in cartesian coordinates. The present work modified the set of anelastic equations for spherical coordinates as shown in equation (\ref{eq0:continuity})-(\ref{eq0:energy}). 

\textit{Continuity Equation}
\begin{equation} \label{eq0:continuity}
\nabla \cdot({\rho}_{a}\textbf{u}) =0
\end{equation}

\textit{Momentum Equation}
\begin{equation} \label{eq0:momentum}
\frac{\partial \textbf{u}}{\partial t} + (\textbf{u}\cdot\nabla)\textbf{u} = -\nabla \frac{p}{\rho_a} + Ra Pr g(r) s e_r + \frac{Pr}{\rho_a} \tau
\end{equation}
 
\textit{Energy Equation}
\begin{equation} \label{eq0:energy}
\frac{\partial s}{\partial t} + (\textbf{u}\cdot\nabla)s = \chi(r) u_r + \frac{1}{\rho_a} \nabla^2 s  + \Phi
\end{equation}
where the velocity vector is expressed as $\textbf{u}$, $t$ is time, $p$ is superadiabatic pressure, $u_r$ is the radial velocity, $s$ is entropy perturbations along the reference state/superadiabatic entropy, $\rho_a$ is the adiabatic density, and $e_r$ is the unit vector in radial direction. The coefficient $\chi(r)$ (in the source term) depends on the radius, which will be discussed later in section \ref{sec:numerical_investigation}. The viscous stress tensor $\tau$ is represented as  
 \begin{equation}
\tau = \nabla ^2 \textbf{u} + \dfrac{1}{3} \nabla(\nabla\cdot \textbf{u}) 
\end{equation}
 and viscous heating term in equation (\ref{eq0:energy}) is expressed as
 \begin{equation}
\Phi=\frac{2\epsilon}{RaT_a\rho_a} \left[e_{ij} -\frac{1}{3}(\nabla\cdot\textbf{u})\delta_{ij}\right]^2
\end{equation}
where $e_{ij} = 1/2 (\partial_j u_i + \partial_i u_j)$ is the strain rate tensor, and $\delta_{ij}$ is the Kronecker delta. The length and time are non-dimensionalized with the domain depths ($l = r_o - r_i$) and $l^2/\kappa$, respectively. The pressure is scaled with $\Delta\rho gl$. The thermal diffusivity $\kappa$ and viscosity $\nu$ are kept constant for simplicity.  The adiabatic temperature $T_a$, adiabatic density $\rho_a$ and gravity $g(r)$ vary along the radial direction only and are time-independent. The expression of $T_a$, $\rho_a$, $g(r)$ and the reference scales employed for non-dimensionalization are given later in section \ref{sec:numerical_investigation}. The non-dimensional control parameters Rayleigh number ($Ra$) and Prandtl number ($Pr$) are presented as $Ra$ $= gl^3 \Delta s/c_p\nu \kappa$ and $Pr$ $=\nu/\kappa$. The Dissipation number is interpreted as $\epsilon = gl/c_p T_a$, where $c_p$ is the specific heat at constant pressure.   

\subsection{Splitting into surface and radial components}\label{sec:split_eq}
To solve the anelastic equations in the spherical shell, the differential operators in equations (\ref{eq0:continuity})-(\ref{eq0:energy}) are split into surface and radial operators as $\Delta = \Delta_\perp + e_r \partial/\partial r$, where subscript $\perp$ indicates the compact Riemannian surface (or spherical surface) and \textit{r} is the radial direction. Similarly, the velocity vector field is split as $\textbf{u} = \textbf{u}_\perp + u_r e_r$, where $u_r$ is the radial velocity along the unit vector $e_r$ and $\textbf{u}_\perp $ is the spherical surface velocity. The velocity on the spherical surface consists of colatitudinal ($\theta$) and longitudinal ($\phi$) direction velocity which is expressed as $\textbf{u}_\perp = u_\theta e_\theta + u_\phi e_\phi$. Therefore, the vector calculus notations of the anelastic equation in spherical surface and radial direction are as follows:

\begin{equation} \label{eq02:continuity}
\nabla_\perp \cdot({\rho}_{a}\textbf{u}_\perp) + \frac{1}{r^2}\frac{\partial}{\partial r} \left(r^2\rho_a u_r\right) =0
\end{equation}
\begin{equation} \label{eq02:momentum}
\begin{split}
\frac{\partial \textbf{u}_\perp}{\partial t} + (\textbf{u}_\perp\cdot\nabla)\textbf{u}_\perp + u_r\frac{\partial \textbf{u}_\perp}{\partial r}+\frac{u_r}{r}\textbf{u}_\perp= -\frac{1}{\rho_a}\nabla_\perp p + \frac{Pr}{\rho_a}\bigg[\nabla_\perp(\nabla_\perp \cdot \textbf{u}_\perp) -\Delta^{R}_\perp \textbf{u}_\perp \\
 + \frac{2\textbf{u}_\perp}{r^2}+\frac{2}{r}\nabla_\perp u_r + \frac{2}{r}\frac{\partial \textbf{u}_\perp}{\partial r} + \frac{\partial^{2}\textbf{u}_\perp}{\partial r^{2}}\bigg] + \frac{Pr}{3 \rho_a}\left[\nabla_\perp(\nabla_\perp\cdot\textbf{u}_\perp) + \nabla_\perp(\frac{\partial u_r}{\partial r}+\frac{2}{r}u_r)\right]
\end{split}
\end{equation}
\begin{equation} \label{eq02:R-momentum}
\begin{split}
\frac{\partial u_r}{\partial t} + (\textbf{u}_\perp\cdot\nabla)u_r - \frac{|\nabla \textbf{u}_\perp|^2}{r}+ u_r\frac{\partial u_r}{\partial r}= -\frac{\partial}{\partial r}\left(\frac{p}{\rho_a}\right) + RaPrg(r)s+ \frac{Pr}{\rho_a}\bigg[\nabla^{2}_{\perp} u_r -\frac{4}{r}\nabla_\perp\cdot\textbf{u}_\perp \\
-\frac{6}{r^2}u_r+\frac{\partial^2 u_r}{\partial r^2}\bigg] + \frac{Pr}{3\rho_a}\left[\frac{\partial^2 u_r}{\partial r^2}-\frac{2}{r^2}u_r+\frac{2}{r}\frac{\partial u_r}{\partial r}-\frac{1}{r}\nabla_\perp\cdot\textbf{u}_\perp+\nabla_\perp\cdot\left(\frac{\partial \textbf{u}_\perp}{\partial r}\right)\right]
\end{split}
\end{equation}
\begin{equation} \label{eq02:energy}
\begin{split}
\frac{\partial s}{\partial t} + (\textbf{u}_\perp\cdot\nabla)s + u_r\frac{\partial s}{\partial r} = \chi(r) u_r + \frac{1}{\rho_a} \left(\nabla^{2}_{\perp} s + \frac{\partial^2 s}{\partial r^2} +\frac{2}{r}\frac{\partial s}{\partial r}\right) + \Phi
\end{split}
\end{equation}
where $\Delta^{R}_\perp$ and $2\textbf{u}_\perp/r^2$ are the surface Laplace-DeRham operator and Gaussian surface curvature term, respectively. The $\Delta^{R}_\perp$ and Gaussian operator are the results of the deformation tensor and the non-commutativity of the second covariant derivative in curved spaces. The vector notation and discretization of $\Phi$ is provided in appendix \hyperref[appendixA]{A}.
\subsection{Expression of the anelastic equations in exterior calculus notation}\label{sec:expression}
Equation (\ref{eq02:continuity})-(\ref{eq02:energy}) are first expressed in exterior calculus notation before discretization. The details about the exterior calculus notations are reported in previous studies \cite{jagad2021primitive, mohamed2016discrete}. Following the work of Mantravadi et al. \cite{mantravadi2023hybrid} for the Boussinesq equation in spherical shells (see Section 2 of Mantravadi et al. \cite{mantravadi2023hybrid} for more details of exterior calculus), the anelastic equations with exterior calculus notation are as follows:
\begin{equation} \label{eq3:continuity}
*d*({\rho}_{a}\wedge \textbf{u}_{\perp}^{\flat}) + \frac{1}{r^2}\frac{\partial}{\partial r} \left(r^2\rho_a \wedge u_r\right) =0
\end{equation}
\begin{equation} \label{eq3:s-momentum}
\begin{split}
\frac{\partial \textbf{u}_{\perp}^{\flat}}{\partial t} + * \textbf{u}_{\perp}^{\flat}\wedge*d\textbf{u}_{\perp}^{\flat} +\frac{1}{2}d(\textbf{u}_\perp\cdot\textbf{u}_\perp)^\flat+ u_r\wedge\frac{\partial \textbf{u}_{\perp}^{\flat}}{\partial r}+\frac{u_r}{r}\wedge \textbf{u}_{\perp}^{\flat} = - \frac{1}{\rho_a} d p + \frac{Pr}{\rho_a} \bigg[d*d*\textbf{u}_\perp^{\flat}  \\
+ *d*d\textbf{u}^{\flat}_{\perp}+ \frac{2\textbf{u}^{\flat}_\perp}{r^2} + \frac{2}{r}d u_r + \frac{2}{r}\frac{\partial \textbf{u}^{\flat}_\perp}{\partial r} + \frac{\partial^{2}\textbf{u}^{\flat}_\perp}{\partial r^{2}}\bigg] +  \frac{Pr}{3\rho_a}\left[d*d*\textbf{u}^\flat_\perp + d(\frac{\partial u_r}{\partial r}+\frac{2}{r}u_{r})\right]
\end{split}
\end{equation}
\begin{equation} \label{eq3:r-momentum}
\begin{split}
\frac{\partial u_r}{\partial t} + *d*(\textbf{u}^\flat_\perp \wedge u_r) - u_r\wedge(*d*\textbf{u}^\flat_\perp) - \frac{(\textbf{u}_\perp \cdot \textbf{u}_\perp)^\flat}{r}+ u_r\wedge\frac{\partial u_r}{\partial r} = - \frac{1}{\rho_a}\frac{\partial p}{\partial r}  \\
+ p\frac{\partial}{\partial r}\left(\frac{1}{\rho^2_a}\right)
+ RaPrg(r)s + \frac{Pr}{\rho_a}\bigg[*d*d u_r -\frac{4}{r}(*d*\textbf{u}^\flat_\perp) -\frac{6}{r^2}u_r+\frac{\partial^2 u_r}{\partial r^2}\bigg]\\
+  \frac{Pr}{3\rho_a}\bigg[\frac{\partial^2 u_r}{\partial r^2}-\frac{2}{r^2}u_r
+\frac{2}{r}\frac{\partial u_r}{\partial r}-\frac{1}{r}*d*\textbf{u}^\flat_\perp+*d*\left(\frac{\partial \textbf{u}^\flat_\perp}{\partial r}\right)\bigg]
\end{split}
\end{equation}
\begin{equation} \label{eq3:energy}
\begin{split}
\frac{\partial s}{\partial t} + *d*(\textbf{u}^{\flat}_\perp\wedge s) - s \wedge(*d*\textbf{u}^{\flat}_\perp)+ u_r\wedge \frac{\partial s}{\partial r} = \chi(r) u_r + \frac{1}{\rho_a} \bigg(*d*d s \\
+ \frac{\partial^2 s}{\partial r^2} +\frac{2}{r}\frac{\partial s}{\partial r}\bigg)
+ \Phi
\end{split}
\end{equation}
where $\flat$ is the flat operator, $d$ is the exterior derivative, $\wedge$ is the wedge product operator and $*$ is the Hodge star operator. The flat operator $\flat$ is used to transform the vector fields to differential forms. The surface velocity vector $\textbf{u}_\perp$ is transformed to the 1-form velocity $\textbf{u}^\flat_\perp$. While radial velocity, pressure, entropy and their radial derivatives are the zero-form variables. 

\section{Discretization}\label{sec:discretization}
The discretization consists of domain discretization and discretization of anelastic equations (\ref{eq3:continuity})-(\ref{eq3:energy}) using DEC-FD in hybrid form. The domain discretization approach is similar to Mantravadi et al. \cite{mantravadi2023hybrid}, while the discretization of the anelastic equation using the DEC-FD method differs and is more complicated.

\subsection{Domain discretization}\label{sec:domain_discretization}
\begin{figure}[htb!]
		\centering
			\subfloat[]{\includegraphics[height=3.2in]{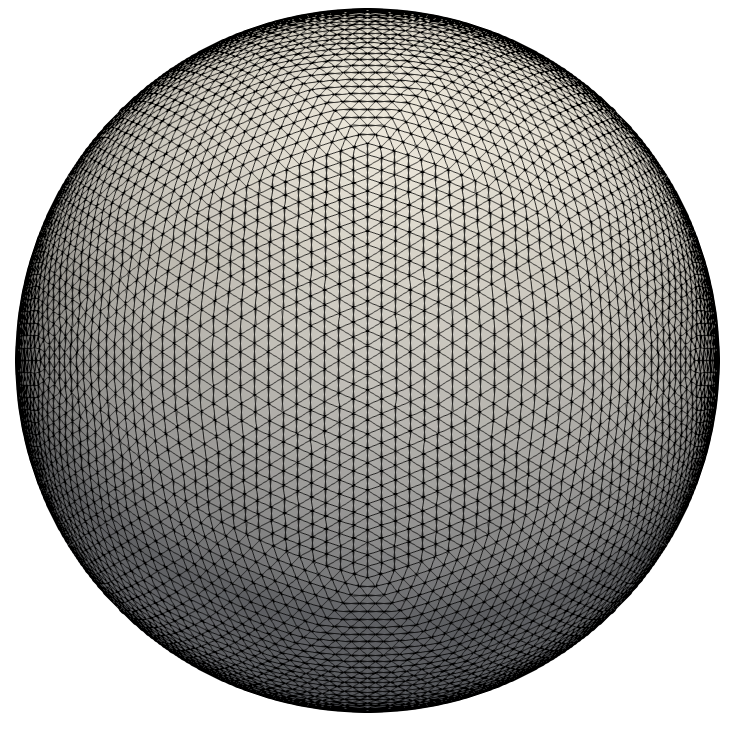}}
            \subfloat[]{\includegraphics[height=3.2in]{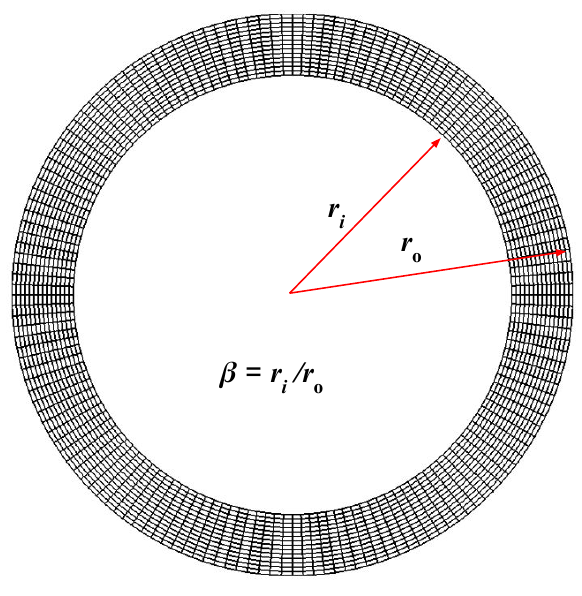}}
   \caption{Computational domain of the spherical shells (a) outer most layer showing prismatic mesh (b) meridional sectional view, where $r_i$ is the inner radius, $r_o$ is the outer radius of the spherical shells and $\beta$ is the aspect ratio.}\label{fig:geometry}
\end{figure}

The inner base spherical surface of radius $r_i$ is first constructed using an icosahedral grid \cite{wang2011geometric}. This base inner spherical surface is extruded in the radial direction to obtain a prismatic mesh, as shown in Figure \ref{fig:geometry}. 
The resolution of the computational domain is maintained with the number of triangular elements on the spherical surface and the number of spherical layers \textit{k} in the radial direction. The simplicial complexes are considered on the spherical surfaces. Based on algebraic topology, the simplicial complexes employed on the spherical surfaces are 2-complexes like nodes, edges, and cells.
Let $\Omega_i$ be the two-dimensional spherical surface physical domain (\textit{N}=2), where subscript \textit{i} represents the $i^{th}$ layer. The simplicial complex $K_i$ has been used to approximate the physical domain $\Omega_i$. The domain simplex is denoted by $\sigma^j_i \in K_i$, where \textit{j} is the dimension; for details on simplices, see previous work \cite{hirani2003discrete,hirani2015numerical,munkres2018elements}. The simplex $\sigma^j_i$ is defined by the nodes forming it. Each primal simplicial complex $K_i$ has a corresponding dual complex $\star K_i$. Dual of the primal simplex $\sigma^j_i$ is $\star\sigma^j_i \in \star K_i$ and it is represented as (N-j)-cell. We have considered circumferential dual mesh, which can maintain a simple expression of the Hodge star operator, leading to the formation of diagonal matrices. In other words, the simplicial mesh consists of primal simplices and corresponding dual cells; for more details, see Figure 1 of Mohamed et al. \cite{mohamed2016discrete}. The circumcenter of a triangle represents a dual of a triangle. The dual of a primal edge corresponds to the dual edge, which connects two adjacent triangles circumcenter and supports the primal edge. The dual of a primal node corresponds to polygons created by the edges dual to the primal edges connected to the primal node. The orientation of primal 2-simplices is either counterclockwise or clockwise. The primal and dual 1-simplices are oriented arbitrarily. The dual 2-simplices follow the orientation of the primal 2-simplices. The primal and dual simplices orientations are consistent on all the spherical layers. Note that we use Delaunay meshes for the simplicial meshes and do not store the dual mesh explicitly.

\subsection{Discretization of anelastic equation}\label{sec:discretization_anelastic}
The brief overview of DEC for two-dimension is reported in previous studies\cite{mantravadi2023hybrid, jagad2021primitive, mohamed2016discrete, hirani2015numerical}. The discretization of anelastic equations using DEC-FD method is presented in this subsection. The exterior calculus operators (or flow on the spherical surface) are discretized using discrete exterior calculus, while the radial differential operators are treated with the finite difference method.
\begin{equation} \label{eq4:continuity}
*_{2,i} d_1 *^{-1}_1 \rho_{a,i}U_{\perp,i} + \frac{1}{r_i^2} D_1 (r_i^2 \rho_{a,i} U_{r,i}) =0
\end{equation}
\begin{equation} \label{eq4:momentum0}
\begin{split}
\frac{\partial U_{\perp,i}}{\partial t} + *_{1} V_{\perp,i} \wedge *_{0,i}^{-1} [-d^T_0]U_{\perp,i} +\frac{1}{2}d^T_1 K_i + U_{r,i}\wedge D_1U_{\perp,i}+\frac{U_{r,i}}{r_i}\wedge U_{\perp,i} = - \frac{1}{\rho_{a,i}}d^T_1 P_i \\
+ \frac{Pr}{\rho_{a,i}}\bigg[d^T_1*_{2,i}d_1*^{-1}_{1}U_{\perp,i}  + *_1 d_0 *^{-1}_{0,i} [-d^{T}_0] U_{\perp,i}+ \frac{2U_{\perp,i}}{r_i^2} + \frac{2}{r_i}d^{T}_1 U_{r,i} + \frac{2}{r_i}D_1 U_{\perp,i} + D_2U_{\perp,i} \bigg] \\
+ \frac{Pr}{3\rho_{a,i}}\left[d^T_1*_{2,i}d_1*^{-1}_{1}U_{\perp,i} + d^T_{1}(D_1U_{r,i}+\frac{2}{r_i}U_{r,i})\right]
\end{split}
\end{equation}
\begin{equation} \label{eq4:momentum}
\begin{split}
\frac{\partial U_{r,i}}{\partial t} + *_{2,i}d_1*^{-1}_1(U_{\perp,i}\wedge U_{r,i}) - U_{r,i}\wedge(*_{2,i}d_1*^{-1}_1U_{\perp,i}) - \frac{K_i}{r_i}+ U_{r,i}\wedge D_1U_{r,i} = - \frac{1}{\rho_{a,i}}D_1P_{i} \\
+ P_iD_1\left(\frac{1}{\rho^2_{a,i}}\right) + RaPrg(r)S_{i} + \frac{Pr}{\rho_{a,i}} \bigg[*_{2,i}d_1*_1d^T_1 U_{r,i} -\frac{4}{r_i}(*_{2,i}d_1*^{-1}_1U_{\perp,i}) -\frac{6}{r_i^2}U_{r,i}+D_2U_{r,i}\bigg] \\
+ \frac{Pr}{3\rho_{a,i}}\bigg[D_2U_{r,i}-\frac{2}{r_i^2}U_{r,i} +\frac{2}{r_i}D_1U_{r,i}-\frac{1}{r_i}*_{2,i}d_1*^{-1}_1U_{\perp,i}+*_{2,i}d_1*^{-1}_1D_1U_{\perp,i}\bigg]
\end{split}
\end{equation}

\begin{equation} \label{eq4:energy}
\begin{split}
\frac{\partial S_i}{\partial t} + *_{2,i}d_1*^{-1}_1(U_{\perp,i}\wedge S_i) - S_i \wedge(*_{2,i}d_1*^{-1}_1U_{\perp,i})+ U_{r,i}\wedge D_1S_i \\
= \chi_i U_{r,i} + \frac{1}{\rho_{a,i}} \bigg(*_{2,i}d_1*_1d^T_1 S_i + D_2 S_i +\frac{2}{r_i}D_1 S_i\bigg)+\Phi
\end{split}
\end{equation}
where $U_{\perp,i}$ is stored on the dual edges, and it contains dual velocity 1-form. The subscript \textit{i} represents the $i^{th}$ spherical layer. Further, $P_i$, $U_{r,i}$, and $S_{i}$ are stored on the dual nodes and contain respective dual 0-forms \textit{p}, $u_r$ and \textit{s}. The vector $K_i$ is stored on dual nodes and consists of discrete dual 0-forms $(\textbf{u}_\perp\cdot\textbf{u}_\perp)^\flat$. The operator $D_\sigma$ is the derivative of finite difference method, where $\sigma$ is the order of derivative. The vector $V_{\perp,i}$ consist of surface velocity of dual 1-form.
The discrete Hodge star $*$ and exterior derivative \textit{d} operator relate the dual and primal mesh complexes on the discrete \textit{j}-forms space. The discrete exterior derivative operator $d_j$ transfers the primal \textit{j}-form to primal (\textit{j}+1)-form. While the discrete exterior derivative operator $d_{N-j-1}$ transforms the dual j-form to dual (\textit{j}+1)-form. The two-dimensional mesh orientation convention results in a negative sign of $d^{T}_0$. The discrete Hodge star operator $*_j$ transfers information between the dual and primal mesh, where $*_j$ maps primal \textit{j}-forms to dual (\textit{N}-\textit{j})-forms. The inverse Hodge star operator $*^{-1}_j$ maps  dual (\textit{N}-\textit{j})-forms to primal \textit{j}-forms.

We obtain the generalized DEC-FD discretization by simplifying the wedge product using the Hirani \cite{hirani2003discrete} approach. The second term $(U_{\perp,i}\wedge U_{r,i})$ in equation (\ref{eq4:momentum}) consists of wedge product of a dual 1-form ($U_{\perp,i}$) and dual 0-form ($U_{r,i}$). The dual 0-forms are stored on the dual nodes, and the dual 1-forms are stored on the dual edges. Therefore, on each dual edge, the average of dual nodes is performed to get dual 1-form on edges, which is represented as a vector $0.5(|d^T_1|U_{r,i})$. Hence, the second term in element-wise multiplication operation is written as $A_iU_{\perp,i}$, where $A_i$ is $N_1\times N_1$ matrix defined as diag$(0.5|d^T_1|U_{r,i})$. Note diag(.) denotes the diagonal matrix composed of the enclosed vector entries. Similarly, the fourth term ($U_{r,i}\wedge D_1U_{\perp,i}$) in equation (\ref{eq4:momentum0}) is expressed as $A_i D_1 U_{\perp,i}$. The second term $(U_{\perp,i}\wedge S_{i})$ in equation (\ref{eq4:energy}) is expressed as $C_iU_{\perp,i}$, where $C_i$ is a $N_1\times N_1$ matrix defined as diag$(0.5|d^T_1|S_{i})$. Here, $N_1$ is the number of dual edges. The third term $U_{r,i}\wedge(*_2d_1*^{-1}_1U_{\perp,i})$ in equation (\ref{eq4:momentum}) consists of wedge product of $U_{r,i}$ and $*_2d_1*^{-1}_1U_{\perp,i}$, where $*^{-1}_1U_{\perp,i}$ is transformed from the dual 1-form to primal 1-form. Subsequently, $d_1*^{-1}_1U_{\perp,i}$ is transformed from primal 1-form to primal 2-form and $*_2d_1*^{-1}_1U_{\perp,i}$ is transformed from primal 2-form to dual 0-form on the dual nodes. Therefore, the wedge product of $U_{r,i}$ (dual 0-form) and $*_2d_1*^{-1}_1U_{\perp,i}$ (dual 0-form) is expressed as $B_i*_2d_1*^{-1}_1U_{\perp,i}$, where $B_i$ is $N_0\times N_0$ matrix defined as diag$(U_{r,i})$. Here $N_0$ is the number of dual nodes. Similarly, the third term $S_i \wedge(*_{2,i}d_1*^{-1}_1U_{\perp,i})$ in equation (\ref{eq4:energy}) is written as $M_{S,i}*_{2,i}d_1*^{-1}_1U_{\perp,i}$, where $M_{S,i}$ is $N_0\times N_0$ matrix defined as diag$(S_{i})$. Finally, the convective term $*_1 V_{\perp,i} \wedge *_0^{-1} [-d^T_0]U_{\perp,i}$ in equation (\ref{eq4:momentum0}) is further simplified as $*_1 W_{v,i} *_0^{-1} [-d^T_0]U_{\perp,i}$, where $W_{v,i}$ is $N_1\times N_2$ sparse matrix defined as 0.5diag$(V_{\perp,i})|d_0|$. Here $N_2$ is the number of dual cells. The $V_{\perp,i}$ vector consists of velocity 1-form stored on the primal edges with $N_1$ entries. The generalized DEC-FD discretized equations are written as:
\begin{equation} \label{eq5:continuity}
*_{2,i} d_1 *^{-1}_1 \rho_{a,i}U_{\perp,i} + \frac{1}{r_i^2} D_1 (r_i^2 \rho_{a,i} U_{r,i}) =0
\end{equation}
\begin{equation} \label{eq5:momentum}
\begin{split}
\frac{\partial U_{\perp,i}}{\partial t} + *_1 W_{v,i} *_{0,i}^{-1} [-d^T_0]U_{\perp,i} +\frac{1}{2}d^T_1 K_i+ A_iD_1U_{\perp,i}+\frac{1}{r_i}A_iU_{\perp,i} = - \frac{1}{\rho_{a,i}}d^T_1 P_i + \frac{Pr}{\rho_{a,i}} \bigg[d^T_1*_{2,i}d_1*^{-1}_{1}U_{\perp,i}  \\
+ *_1 d_0 *^{-1}_{0,i} [-d^{T}_0] U_{\perp,i}+ \frac{2U_{\perp,i}}{r_i^2} + \frac{2}{r_i}d^{T}_1 U_{r,i} + \frac{2}{r_i}D_1 U_{\perp,i} + D_2U_{\perp,i} \bigg] \\
+ \frac{Pr}{3\rho_{a,i}}\left[d^T_1*_{2,i}d_1*^{-1}_{1}U_{\perp,i} + d^T_{1}(D_1U_{r,i}+\frac{2}{r_i}U_{r,i})\right]
\end{split}
\end{equation}
\begin{equation} \label{eq5:momentum}
\begin{split}
\frac{\partial U_{r,i}}{\partial t} + *_{2,i}d_1*^{-1}_1A_iU_{\perp,i} - B_i*_{2,i}d_1*^{-1}_1U_{\perp,i} - \frac{K_i}{r_i}+ B_iD_1U_{r,i} = - \frac{1}{\rho_{a,i}}D_1P_{i} + P_iD_1\left(\frac{1}{\rho^2_{a,i}}\right)\\
+ RaPrg(r)S_{i} + \frac{Pr}{\rho_{a,i}}\bigg[*_{2,i}d_1*_1d^T_1 U_{r,i} -\frac{4}{r_i}(*_{2,i}d_1*^{-1}_1U_{\perp,i}) -\frac{6}{r_i^2}U_{r,i}+D_2U_{r,i}\bigg]
+ \frac{Pr}{3\rho_{a,i}}\bigg[D_2U_{r,i}-\frac{2}{r_i^2}U_{r,i} \\
+\frac{2}{r_i}D_1U_{r,i}-\frac{1}{r_i}*_{2,i}d_1*^{-1}_1U_{\perp,i}+*_{2,i}d_1*^{-1}_1D_1U_{\perp,i}\bigg]
\end{split}
\end{equation}
\begin{equation} \label{eq5:energy}
\begin{split}
\frac{\partial S_i}{\partial t} + (*_{2,i}d_1*^{-1}_1 C_i - M_{S,i} *_{2,i}d_1*^{-1}_1)U_{\perp,i}+B_i D_1S_i = \chi_i U_{r,i} \\ + \frac{1}{\rho_{a,i}} \bigg[ *_{2,i}d_1*_1d^T_1 S_i + D_2 S_i + \frac{2}{r_i}D_1 S_i  \bigg] + \Phi
\end{split}
\end{equation}

For radial spatial derivatives, we use a second-order finite difference method. A semi-implicit time integration scheme is employed to evolve the equation in time, where non-linear terms are treated explicitly, and linear terms are implicitly treated. Discrete equations are further simplified as follows:
\begin{equation} \label{eq6:continuity}
*_{2,i} d_1 *^{-1}_1 \rho_{a,i}U^{n+1}_{\perp,i} + \frac{\rho_{a,i+1}U^{n+1}_{r,i+1}}{2\Delta r} - \frac{\rho_{a,i-1}U^{n+1}_{r,i-1}}{2\Delta r} + \frac{2}{r_i} \rho_{a, i}U^{n+1}_{r,i} =0
\end{equation}
\begin{equation} \label{eq6:momentum}
\begin{split}
\left[\left(\frac{1}{\Delta t} - \frac{2Pr}{\rho_{a,i} r_i^2} + \frac{2Pr}{\rho_{a,i} (\Delta r)^2}\right)\mathbb{I} - \frac{4Pr}{3\rho_{a,i}} d^T_1*_{2,i}d_1*^{-1}_{1} - \frac{Pr}{\rho_{a,i}} *_1 d_0 *^{-1}_{0,i} [-d^{T}_0] \right] U^{n+1}_{\perp,i}\\
- \left[\left(\frac{Pr}{\rho_{a,i} r_i\Delta r}+\frac{Pr}{\rho_{a,i}(\Delta r)^2}\right)\mathbb{I}\right]U^{n+1}_{\perp,i+1} + \left[\left(\frac{Pr}{\rho_{a,i} r_i\Delta r}-\frac{Pr}{\rho_{a,i}(\Delta r)^2}\right)\mathbb{I}\right]U^{n+1}_{\perp,i-1}-\frac{8Pr}{3\rho_{a,i} r_i}d^T_1U^{n+1}_{r,i}\\
+\frac{1}{\rho_{a,i}}d^T_1 P^{n+1}_{i} - \frac{Pr}{3\rho_{a,i}} d^T_1 \frac{U^{n+1}_{r,i+1}}{2\Delta r} + \frac{Pr}{3\rho_{a,i}} d^T_1 \frac{U^{n+1}_{r,i-1}}{2\Delta r} = \left(\frac{1}{\Delta t}\mathbb{I} - *_1 W^n_v *_{0,i}^{-1} [-d^T_0] - \frac{1}{r_i}A_i^n \right)U^n_{\perp,i}\\
-\frac{1}{2\Delta r}A_i^n U^n_{\perp,i+1}+\frac{1}{2\Delta r}A_i^n U^n_{\perp,i-1} - \frac{1}{2}d^T_1 K_i^n 
\end{split}
\end{equation}
\begin{equation} \label{eq6:momentum}
\begin{split}
\left[\left(\frac{1}{\Delta t} + \frac{20Pr}{3\rho_{a,i} r_i^2} + \frac{8Pr}{3\rho_{a,i}(\Delta r)^2} \right)\mathbb{J} - \frac{Pr}{\rho_{a,i}} *_{2,i}d_1*^{-1}_{1}d^T_1 \right] U^{n+1}_{r,i} + \frac{4Pr}{\rho_{a,i} r_i}*_{2,i} d_1 *^{-1}_1 U^{n+1}_{\perp,i} 
+ \frac{1}{\rho_{a,i}} \frac{P^{n+1}_{i+1}}{2\Delta r} \\
- \frac{1}{\rho_{a,i}} \frac{P^{n+1}_{i-1}}{2\Delta r} - RaPrg(r)S^{n+1}_{i} + \frac{Pr}{3\rho_{a,i} r_i}*_{2,i} d_1 *^{-1}_1 U^{n+1}_{\perp,i} - \bigg[\bigg(\frac{Pr}{3\rho_{a,i} r_i\Delta r}
+\frac{4Pr}{3\rho_{a,i}(\Delta r)^2}\bigg)\mathbb{J}\bigg]U^{n+1}_{r,i+1} \\
+ \left[\left(\frac{Pr}{3\rho_{a,i} r_i\Delta r}-\frac{4Pr}{3\rho_{a,i}(\Delta r)^2}\right)\mathbb{J}\right]U^{n+1}_{r,i-1}
- \frac{Pr}{3\rho_{a,i}} *_{2,i} d_1 *^{-1}_1 \frac{U^{n+1}_{\perp,i+1}}{2\Delta r} + \frac{Pr}{3\rho_{a,i}} *_{2,i} d_1 *^{-1}_1 \frac{U^{n+1}_{\perp,i-1}}{2\Delta r} \\
= \frac{1}{\Delta t}U^n_{r,i} + \left(- *_{2,i} d_1 *^{-1}_1 A_i^n + *_{2,i} d_1 *^{-1}_1 B_i^n \right)U^n_{\perp,i}
-\frac{1}{2\Delta r}B_i^n U^n_{r,i+1}+\frac{1}{2\Delta r}B_i^n U^n_{r,i-1} \\
+ \frac{1}{r_i} K_i^n + M_{P,i} \frac{\left(\frac{1}{\rho^2_{a}}\right)_{i+1}}{2\Delta r} - M_{P,i} \frac{\left(\frac{1}{\rho^2_{a}}\right)_{i-1}}{2\Delta r}
\end{split}
\end{equation}
\begin{equation} \label{eq6:energy}
\begin{split}
\bigg[\bigg(\frac{1}{\Delta t} + \frac{2}{\rho_{a,i} (\Delta r)^2}\bigg) \mathbb{J} - \frac{1}{\rho_{a,i}} *_{2,i}d_1*_1d^T_1 \bigg] S^{n+1}_{i} - \frac{1}{\rho_{a,i} \Delta r}\bigg(\frac{1}{r_i} + \frac{1}{\Delta r} \bigg)S^{n+1}_{i+1}\\
+\frac{1}{\rho_{a,i} \Delta r}\bigg(\frac{1}{r_i} - \frac{1}{\Delta r}\bigg)S^{n+1}_{i-1} - \chi_iU^{n+1}_{r,i}= \frac{1}{\Delta t} S^{n}_{i} + ( - *_{2,i}d_1*^{-1}_1 C^n_i + M^n_{S,i} *_{2,i}d_1*^{-1}_1)U^n_{\perp,i}\\
- \frac{1}{2\Delta r}B^n_i S^{n}_{i+1} + \frac{1}{2\Delta r}B^n_i S^{n}_{i-1} + \Phi
\end{split}
\end{equation}
where $\Delta r$ is the grid spacing in the radial direction, $\Delta t$ is the time step. $M_{P,i}$ is $N_0\times N_0$ matrix defined as diag$(P_{i})$. $\mathbb{I}$ and $\mathbb{J}$ are the unit matrices with $N_1\times N_1$ and $N_0\times N_0$ entries, respectively. Some information about computation of $\Phi$ is given in appendix \hyperref[appendixA]{A}.
\section{Implementation} \label{sec:implementation}
\begin{figure}[htb!]
    \centering
    \includegraphics[width=1\linewidth]{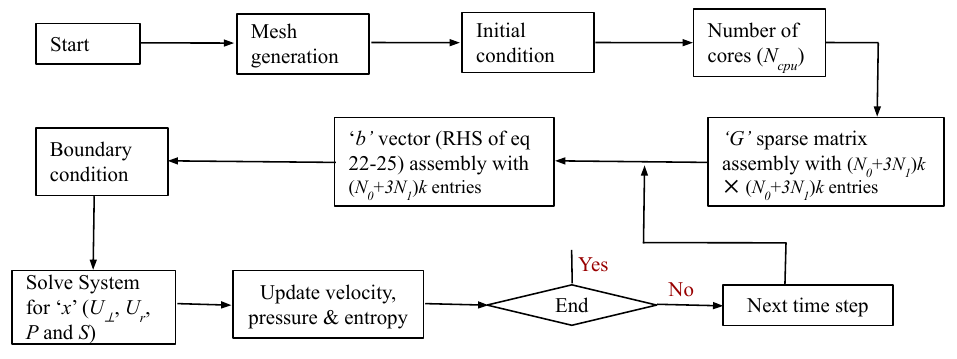}
    \caption{ Solver implementation with global sparse matrix '\textit{G}' and vector '\textit{b}' assembly in parallel.}
    \label{fig:flow}
\end{figure}

The hybrid DEC-FD code is written within a PETSc (Portable and Extensible Toolkit for Scientific Computing) framework for flexible data structures, modular operation, user interference, mesh generation, MPI parallel, and memory allocation.  Figure \ref{fig:flow} illustrates the implementation of a hybrid DEC-FD solver. The DMPlex library within the PETSc framework has been employed to generate the base two-dimensional spherical surface. Subsequently, the base spherical mesh is extruded in radial direction to get discrete three-dimensional spherical shells. The computing nodes $N_{cpu}$ is considered such that $N_{cpu}$=$N_{var}N_{lyr}$, where $N_{var}$ is number of variables ($U_\perp$, $U_r$, \textit{P} and \textit{S}) and $N_{lyr}$ is number of spherical shell layers. This parallel environment helps to balance the load on the nodes, and no single node is overloaded. The discretized DEC-FD anelastic equations (\ref{eq6:continuity})-(\ref{eq6:energy}) are solved as fully coupled system \textit{Gx}=\textit{b}, where \textit{G} is a sparse martix with structure of $(N_1+3N_0)k \times (N_1+3N_0)k$, \textit{b} is a vector with $(N_1+3N_0)k$ entries, and \textit{x} corresponds to a solution vectors ($U_{\perp,1}, U_{\perp,2}, ..., U_{\perp,k-1}$, $U_{r,1}, U_{r,2}, ..., U_{r,k-1}$, $P_{1}, P_{2}, ..., P_{k-1}$, $S_{1}, S_{2}, ..., S_{k-1}$). The global sparse matrix '\textit{G}' and vector '\textit{b}' are assembled in parallel. The boundary conditions for the velocity, pressure and entropy are implemented. A PETSc parallel linear solver known as MUMPS (MUltifrontal Massively Parallel sparse direct Solver) is employed to solve the sparse matrix in parallel.
\section{Numerical Investigation}\label{sec:numerical_investigation}
The anelastic convection of the fluid layer is considered in the non-rotating spherical shell, as shown in Figure \ref{fig:geometry}. The fluid layer is confined in a spherical shell of thickness \textit{l}, and the gravity $\textbf{g}=ge_r$ points towards the center of the sphere along unit vector $e_r$. 
We assume that the gravity is proportional to $1/r^2$, and we use the same background state as in the anelastic benchmark paper of Jones et al. \cite{jones2011anelastic}. The background reference states are expressed as
\begin{equation}
\rho_a=\zeta^n, \hspace{10pt}T_a=\zeta, \hspace{10pt}P_a=\zeta^{n+1}, \hspace{10pt}\zeta=c_0 + c_1l/r
\end{equation}
where \textit{n} is the polytropic index, $l=r_o - r_i$ and the constants $c_0$ and $c_1$ are defined by\\
\vspace{5pt}
$\zeta_o=\frac{\beta + 1}{\beta exp(N_{\rho}/n) + 1}$, $\zeta_i = \frac{1+\beta-\zeta_o}{\beta}$\\
\vspace{5pt}
$c_0 = \frac{2 \zeta_o - \beta -1}{1-\beta}$, $c_1 = \frac{(1+\beta)(1-\zeta_o)}{(1-\beta)^2}$ \\
\vspace{5pt}
where $\zeta_i$ and $\zeta_o$ are the values of $\zeta$ on the inner and outer boundaries, and $N_\rho$ is the number of density scale across the shell. Following Jones et al. \cite{jones2011anelastic}, the temperature ($T_a$) and density ($\rho_a$) are scaled with respective reference temperature ($T_c$) and density ($\rho_c$) i.e. midway between the inner and outer layer. The gravity $g(r)$ is considered proportional to $1/r^2$ and scaled with a reference point midway between the inner and outer layers. The number of density scale height across the layer is defined as $N_\rho=\ln (\rho_a(r_i)/\rho_a(r_o))$.
 
The coefficient $\chi(r)$ in equation (\ref{eq0:energy}) is used to mimic the convection and it is defined as $\partial S_a / \partial r$, where $S_a$ is the entropy profile. We have chosen the entropy profile $S_a$ for the spherical shells as a function of radius proposed by Gastine \cite{gastine2015turbulent}. 
\begin{equation} \label{eq1:energy}
S_a=\frac{\beta}{(1-\beta)^2}\frac{1}{r} - \frac{\beta}{1-\beta} 
\end{equation}
where $\beta = r_i/r_o$ is the aspect ratio. This model (equation (\ref{eq1:energy})) has the advantage of being independent of the convective flow field and background stratification. The first-order background entropy gradient $\partial S_a/\partial r$ creates an unstable convective atmosphere (see Glatzmaier \cite{glatzmaier1984numerical}). Later, when the convection develops, the entropy field in the domain depends on the \textit{S(t)}. According to previous studies \cite{o2016velocity,hotta2014high}, the flow characteristics are insensitive to how the background entropy gradient is modeled for both rotating and non-rotating simulations. No-slip boundary conditions are imposed on the inner and outer spherical shells. The simulations are carried out at various Rayleigh numbers and density scale height $N_\rho$ = 0-2. The Prandtl number is considered \textit{Pr}=1. The polytropic index is defined as $n=1/(\gamma -1)$, where the specific heat ratio $\gamma$ is set to $5/3$. The aspect ratio of the spherical shells is considered as $\beta$=0.759.

Using four different perspectives, we verify and validate our hybrid discrete exterior calculus and finite difference solver for anelastic thermal convection in spherical shells. In section \ref{sec:MMS}, we perform the code verification using the method of manufactured solution. In section \ref{sec:validation}, we check the accuracy of our solver for thermal convection in spherical shells. In section \ref{sec:structures}, we look in detail at the formation of convective flow structures on the spherical layer. In contrast, in section \ref{sec:energy}, quantitative analysis of energy dynamics is demonstrated. Further, in section \ref{sec:density}, the effect of density ratio on convection is performed.

\subsection{Verification of solver using the Method of Manufactured Solutions - MMS} \label{sec:MMS}
The Method of Manufactured Solutions (MMS) is becoming a conventional and classical method to verify the numerical code. MMS is an accepted approach \cite{waltz2014manufactured,negi2021train,gfrerer2018code,choudhary2016code} to quantify the numerical methods for complex physical systems. In the MMS, the solution is manufactured prior, and the manufactured solution (MS) is substituted in the governing equation to obtain the residual (or source term). The manufactured solution can be physically unrealistic. We consider the manufactured solution as sin and cos function for $\bm{\widetilde{u}} (\tilde u_r, \tilde u_\theta, \tilde u_\phi)$, $\tilde p$, and $\tilde s$ of the form
\begin{equation} \label{eq:MS-ur}
\tilde u_r = K_1 e^{-t} \sin (r) \cos (\theta) \cos (\phi)
\end{equation}
\begin{equation} \label{eq:MS-utheta}
\tilde u_\theta = e^{-t} \cos (r) \cos (\theta) \cos (\phi)
\end{equation}
\begin{equation} \label{eq:MS-uphi}
\tilde u_\phi = 1
\end{equation}
\begin{equation} \label{eq:MS-P}
\tilde p = e^{-t} (\cos (r) + \cos (\theta) + \cos (\phi)) 
\end{equation}
\begin{equation} \label{eq:MS-S}
\tilde s = K_2 e^{-t} (\cos (r) + \cos (\theta) + \cos (\phi))  
\end{equation}
where $K_1$ and $K_2$ are constant parameters. We substitute the MS in the governing equations (\ref{eq0:continuity})-(\ref{eq0:energy}). Since the MS is not the original solution, we get residue terms $R_c$, $R_m$, and $R_s$ for respective continuity, momentum, and energy equations. The residue terms are introduced in the governing equations as a source term and are reported in appendix \hyperref[appendixB]{B}. The residue (or source) terms are generated using the python (sympy \cite{meurer2017sympy}) package.  The symbolic packages are recommended\cite{salari2000code,negi2021train} to generate source terms to avoid mistakes. The modified governing equations are written as
\begin{equation} \label{mms:continuity}
\nabla \cdot({\rho}_{a}\textbf{u}) = R_c
\end{equation}
\begin{equation} \label{mms:momentum}
\frac{\partial \textbf{u}}{\partial t} + (\textbf{u}\cdot\nabla)\textbf{u} = -\nabla \frac{p}{\rho_a} + Ra Pr g(r) s e_r + \frac{Pr}{\rho_a}\left(\nabla ^2 \textbf{u} + \dfrac{1}{3} \nabla(\nabla\cdot \textbf{u})\right)  + R_m
\end{equation}
\begin{equation} \label{mms:energy}
\frac{\partial s}{\partial t} + (\textbf{u}\cdot\nabla)s = \chi(r) u_r + \frac{1}{\rho_a} \nabla^2 s  + \frac{2\epsilon}{RaT_a\rho_a} \left[e_{ij} -\frac{1}{3}(\nabla\cdot\textbf{u})\delta_{ij}\right]^2 + R_s
\end{equation}
Finally, the modified governing equations are solved using a hybrid DEC-FD solver. MS is used to implement the initial condition and boundary condition.
The velocity and entropy values were obtained by solving equations (\ref{mms:continuity})-(\ref{mms:energy}). The difference between the numerical solution and the manufactured solution at a discrete point in the spherical shell is defined as discretization error, and it is evaluated using $L_2$ norm as
\begin{equation}\label{eq:lnorm}
    \Psi^h = L_2 = \sqrt{\frac{1}{N_n}\sum_{r,\theta,\phi} \alpha^D_{(r,\theta,\phi)}-\alpha^M_{(r,\theta,\phi)}}
\end{equation}
where $\alpha^D$ represents the discrete value of the variable of interest obtained from the solution of the modified governing equations (\ref{mms:continuity})-(\ref{mms:energy}), in contrast, $\alpha^M$ is the manufactured solution of the corresponding variable evaluated at a discrete point in spherical shells. $N_n$ is the number of grid points.

\begin{figure}[htb!]
		\centering
			\includegraphics[height=4in]{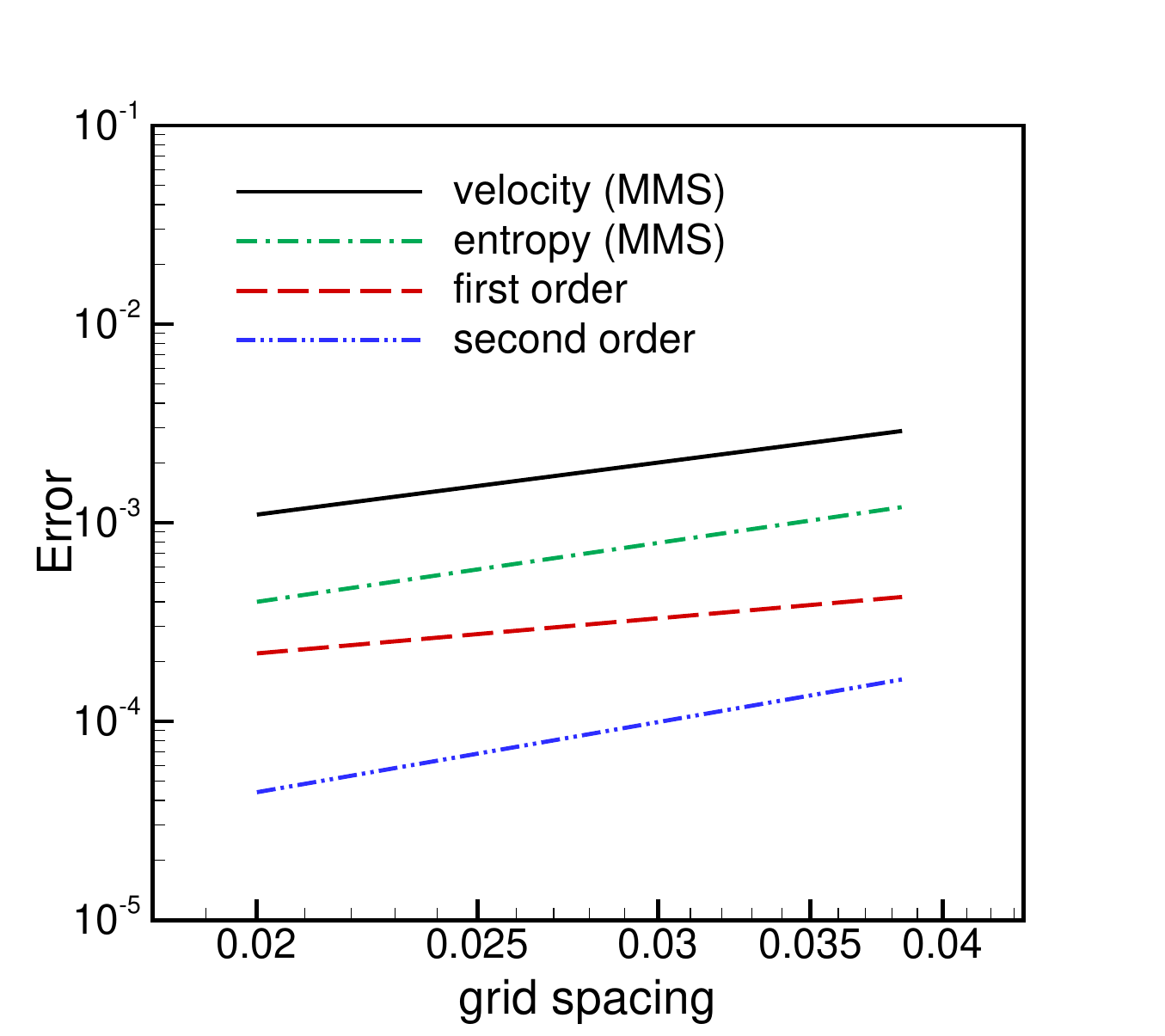}
        \caption{The discretization error order of accuracy at different grid resolution.} \label{fig:mms1}
\end{figure}

The order of accuracy test is the most rigorous acceptance criterion to verify a code. This test involves a systematic grid convergence test by performing a series of numerical simulations. The discretization error is calculated for each grid resolution from equation (\ref{eq:lnorm}). Next, we calculate the error ratio $E_r = log(\Psi^{2h}/\Psi^{h})$, where $\Psi^{2h}$ and  $\Psi^{h}$ are calculated at respective grid sizes 2\textit{h} and \textit{h}. The order of accuracy is then calculated as $OOA = \frac{E_r}{log(b)}$, where \textit{b} is the grid refinement ratio. Figure \ref{fig:mms1} illustrates the discretization error order of accuracy at different grid resolutions. It is observed that the order of accuracy for both velocity and entropy is close to second order. Figure \ref{fig:mms1} confirms that the order of accuracy of DEC-FD discretized equations is in good agreement with the theoretical order of accuracy. The MMS test verifies that the numerical model used in the present work is accurate.

\subsection{Hybrid DEC-FD solver accuracy considerations} \label{sec:validation}
\begin{figure}[htb!]
		\centering
			\includegraphics[height=3.4in]{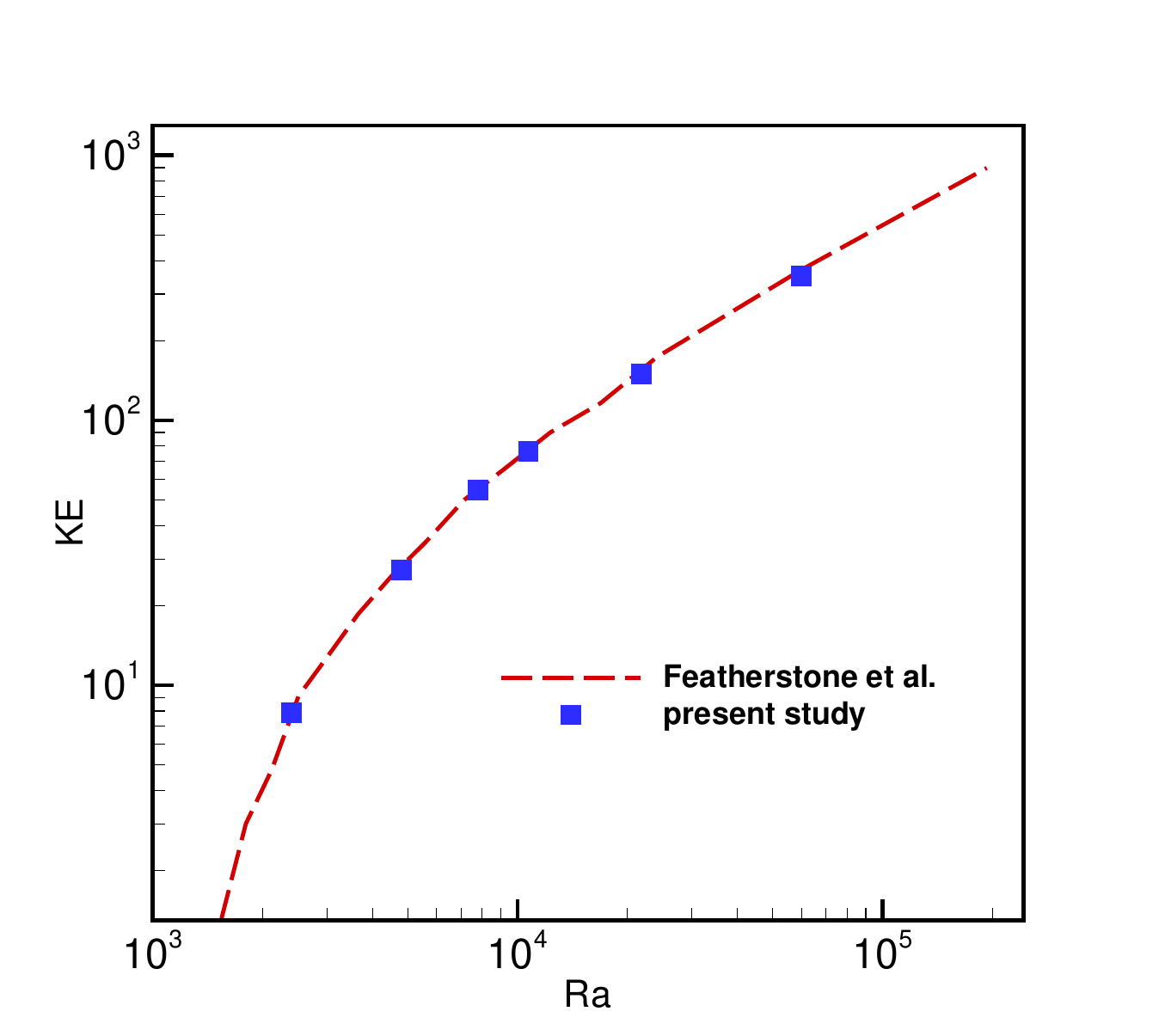}
            \caption{The average kinetic energy for different $Ra$ at $N_\rho$ = 2. The solid squares are the present results which are in good agreement with the results of Featherstone et al. \cite{featherstone2016spectral}(dashed lines).} \label{fig:validate-ke}
	\end{figure}

\begin{figure}[htb!]
		\centering
			\includegraphics[height=3.4in]{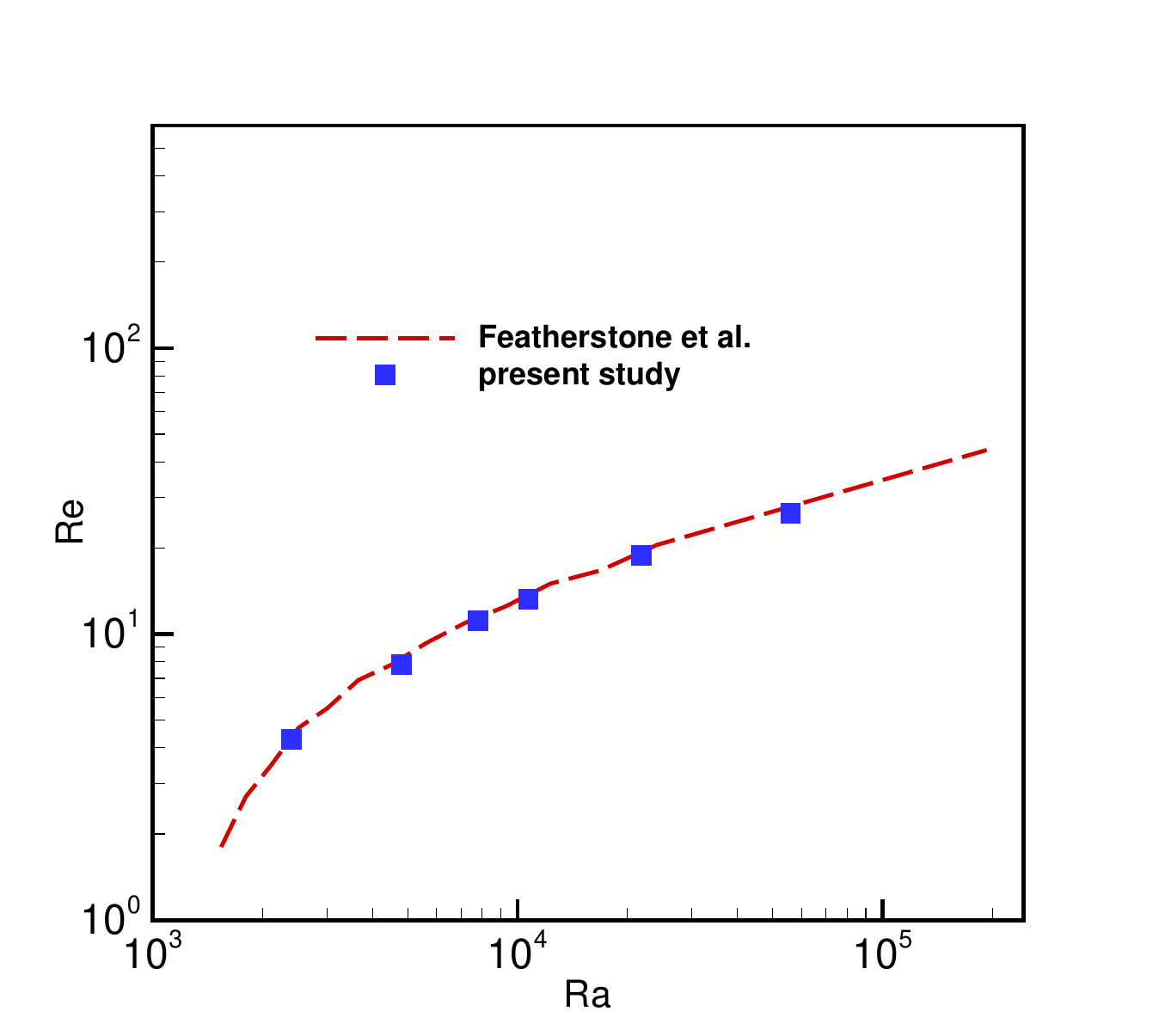}
            \caption{Reynolds number for different $Ra$ at $N_\rho$ = 2. The solid squares are the present results which are in good agreement with the results of Featherstone et al. \cite{featherstone2016spectral}(dashed lines).} \label{fig:validate-re}
\end{figure}

Anelastic equations are being discretized and solved for the first time using the hybrid DEC-FD method. The accuracy of our hybrid DEC-FD solver for anelastic convection in spherical shells is now verified by comparing the results with other benchmark results. We conducted simulations based on the benchmark paper \cite{featherstone2016spectral} for Prandtl number \textit{Pr}=1, density scale height $N_\rho$ = 2, and aspect ratio $\beta$=0.759 at moderate Rayleigh numbers. In order to examine the solver, convective kinetic energy is calculated. The kinetic energy is defined as KE$ = 1/2\langle \rho_a \textbf{u}^2 \rangle$; where the bracket $\langle\cdots\rangle$ represents the temporal and volume average. Figure \ref{fig:validate-ke} shows the kinetic energy as a function of the Rayleigh number for the current model and the benchmark model, where the curve illustrates that the KE and $Ra$ follow a power-law relationship. It is observed that the present results are in good agreement with the benchmark model. In addition, we calculated Reynolds number, which is defined as Re$=\langle \sqrt{\langle u_r^2 + \textbf{u}_\perp^2 \rangle_v} \rangle_t$. Figure \ref{fig:validate-re} presents the Reynolds number with respect to the Rayleigh number for the current and benchmark model. As can be seen, the results presented here are reasonably consistent with those of the benchmark model.
\begin{figure}[htb!]
		\centering
			\includegraphics[height=3.4in]{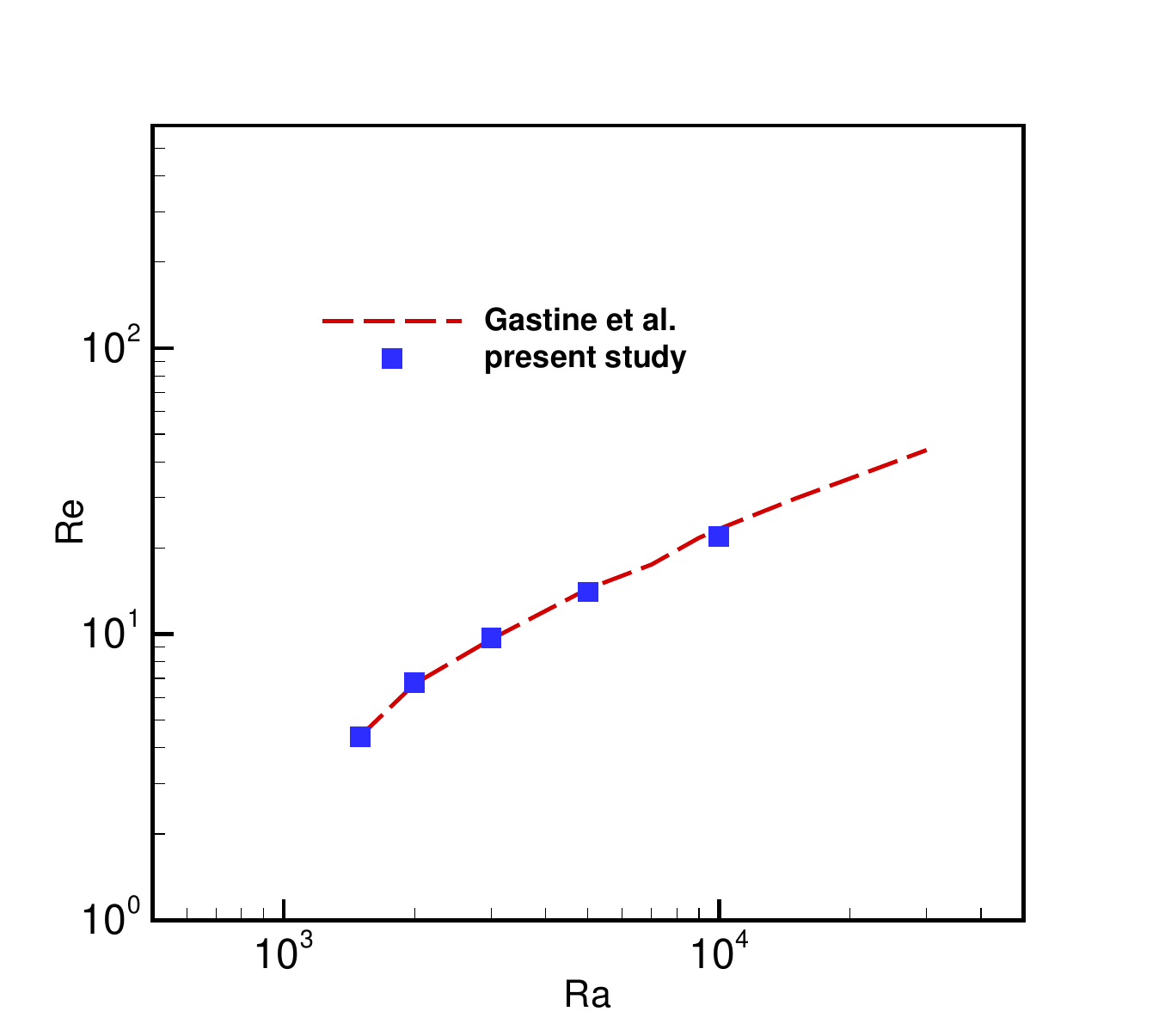}
            \caption{Reynolds number for different $Ra$ at $N_\rho$ = 0. The solid squares are the present results which are in good agreement with the results of Gastine et al. \cite{gastine2015turbulent}(dashed lines).} \label{fig:validate-n0}
\end{figure}

To further test the accuracy of our hybrid solver, we solved anelastic equations (\ref{eq6:continuity})-(\ref{eq6:energy}) for density scale height $N_\rho = 0$, Prandtl number \textit{Pr}=1, aspect ratio $\beta$=0.6 at moderate Rayleigh number. We compared the present results for $N_\rho = 0$ with the benchmark results \cite{gastine2015turbulent}. Figure \ref{fig:validate-n0} shows the Reynolds number of the convective flow as a function of the Rayleigh number for the present work and the benchmark paper. Note that the governing equation used for this case ($N_\rho = 0$) is different from the benchmark model\cite{gastine2015turbulent} governing equation. Indeed, with the control of background reference state ($T_a$ and $\rho_a$), the present solver is able to capture the flow dynamics for both with density ($N_\rho > 0$) and without density ($N_\rho = 0$) stratification. Figures \ref{fig:validate-ke}-\ref{fig:validate-n0} illustrate that the results obtained from the present solver are accurate at $N_\rho = 0$ and $N_\rho > 0$. Continuing to benchmark our hybrid DEC-FD solver for anelastic thermal convection in spherical shells, in succeeding sections, we examined the overall structures, energy dynamics, and density ratio using Prandtl number \textit{Pr}=1, density scale height $N_\rho = 0-2$, aspect ratio $\beta$=0.759 at moderate Rayleigh numbers.

\subsection{Overall structures} \label{sec:structures}
\begin{figure}[htb!]
		\centering
			\subfloat[]{\includegraphics[height=2.0in]{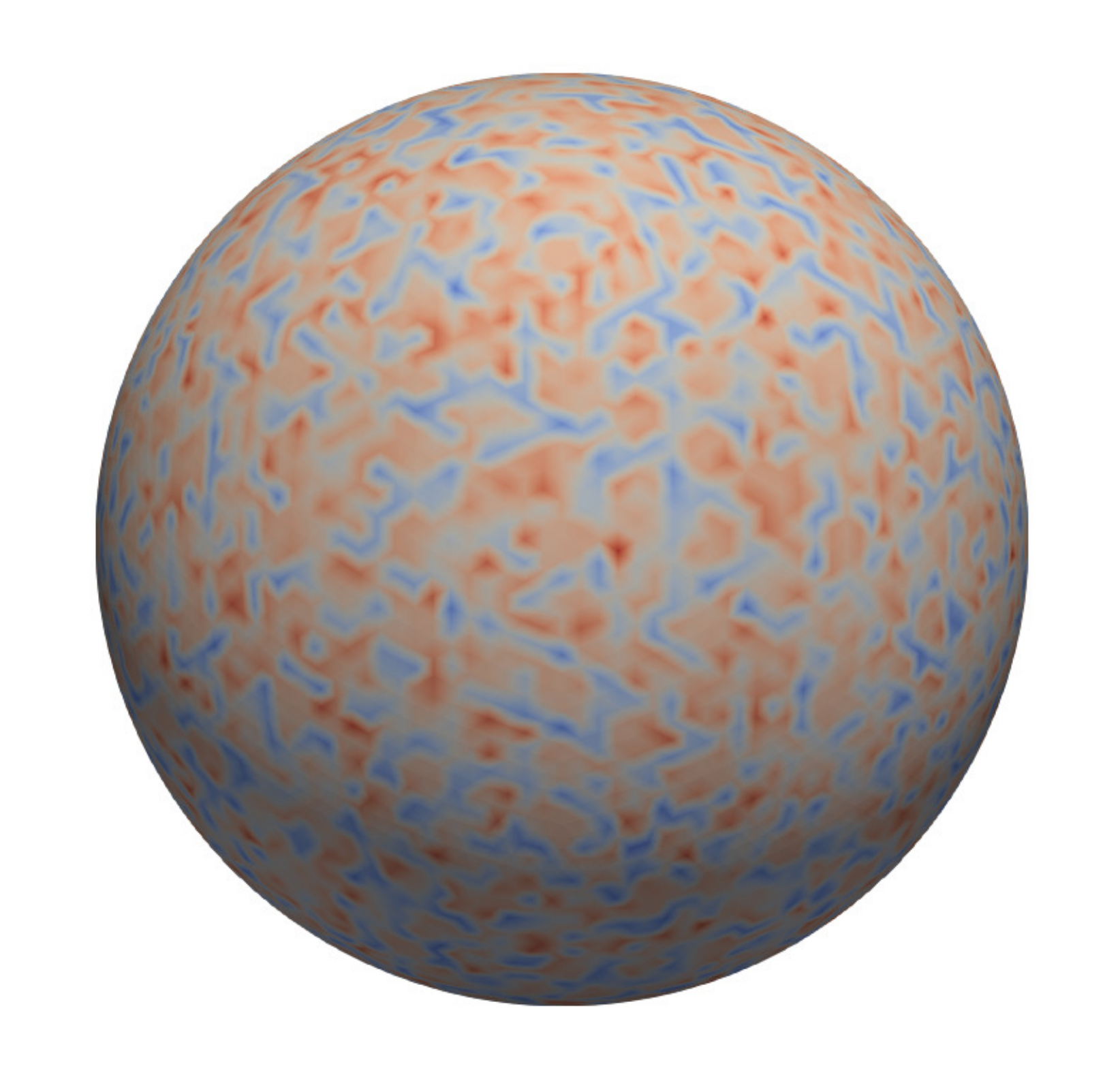}}
            \subfloat[]{\includegraphics[height=2.0in]{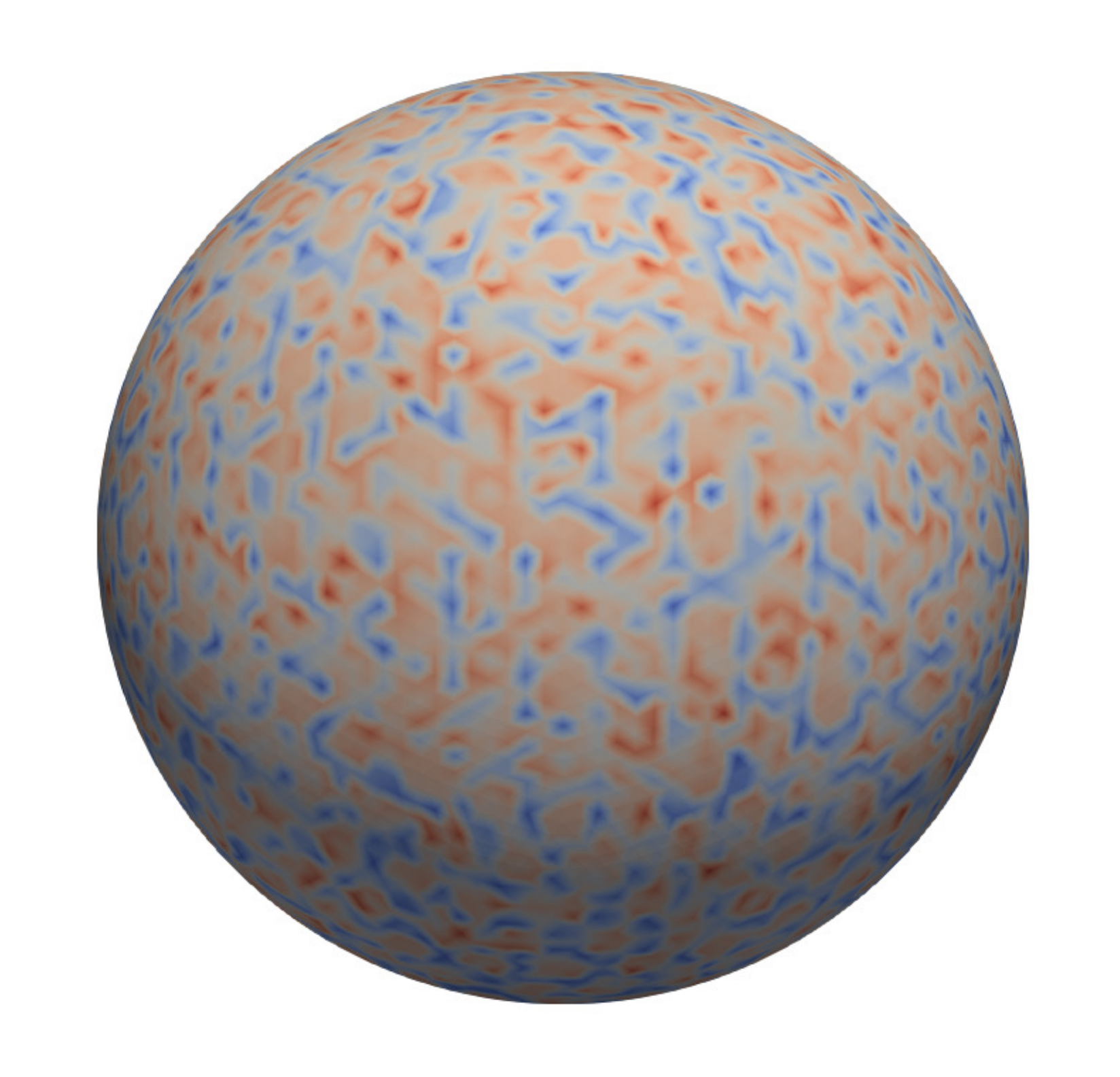}}
            \subfloat[]{\includegraphics[height=2.0in]{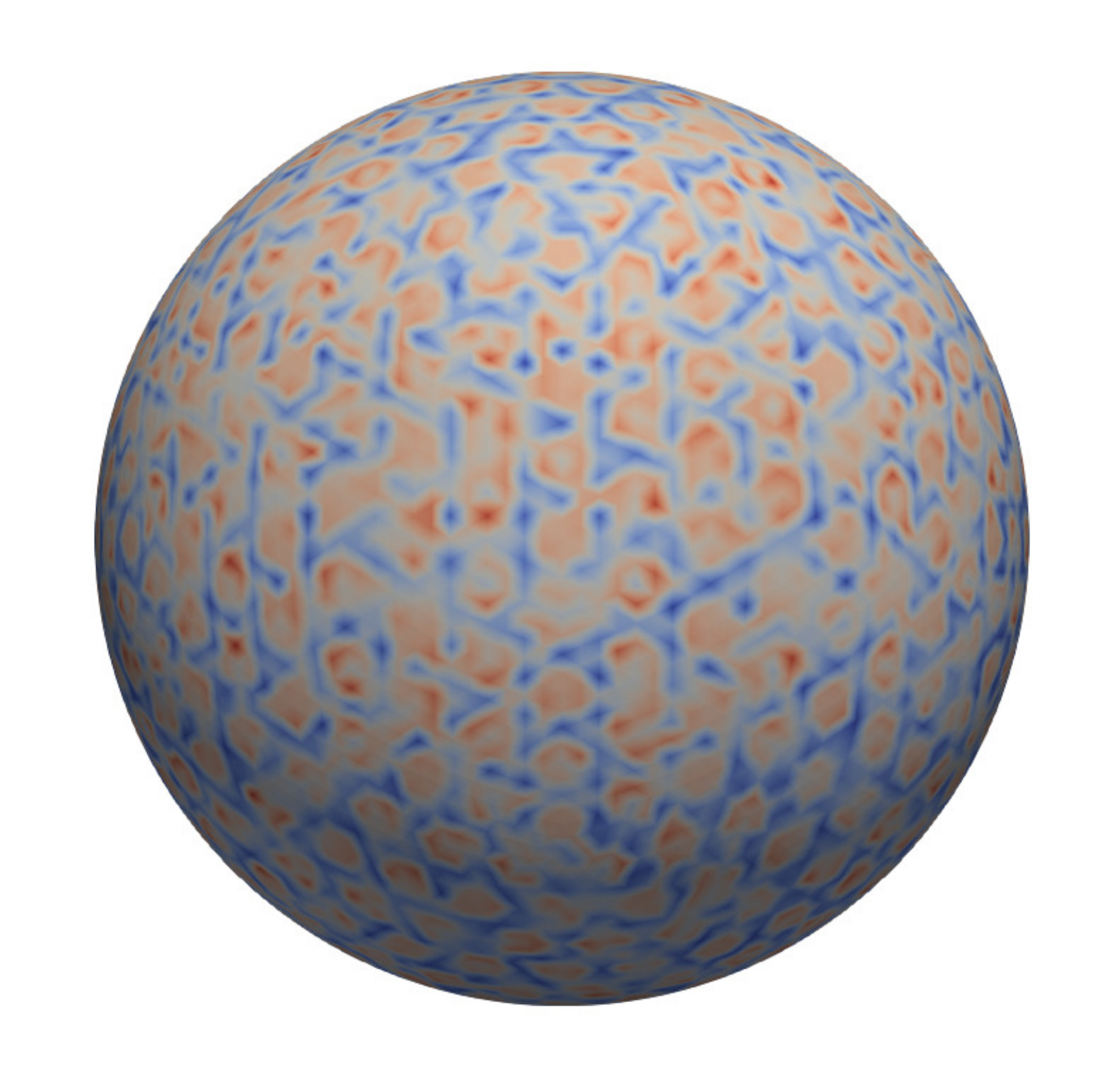}}
            \caption{Entropy fluctuations at the outer layer of the spherical shells at $N_\rho$ = 2 for (a) $Ra$=3.6$\times$10$^3$, (b) $Ra$=5.6$\times$10$^3$, and (c) $Ra$=1.2$\times$10$^4$ (warm plumes in red and cold plumes in blue).} \label{fig:structure-entropy}
\end{figure}

We explore the structures of the convective flow in the spherical shells in this section. Figure \ref{fig:structure-entropy} displays the entropy fluctuation (superadiabatic entropy) of convective flow at the outer layer of the spherical shell at $N_\rho = 2$ for $Ra$=3.6$\times$10$^3$, $Ra$=5.6$\times$10$^3$ and $Ra$=1.2$\times$10$^4$. 
The spherical slice represents a single snapshot in time, where the blue spots represent cool plumes of fluid sinking inwards from the surface, and the red spots are associated with warm-upflows. In general, it is well known that the stratification influences the convective patterns, and more intricate convection are observed at higher Rayleigh number. Indeed, the cold plumes are thought to drive the convection and mixing of the deeper layers of the star. In Figure \ref{fig:structure-entropy}, the entropy fluctuation strength of cold plumes increases from the low $Ra$ to high $Ra$. Compared to the left panel of Figure \ref{fig:structure-entropy}, the right panel shows that cold plume patterns delineate the convective cells. As expected, the entropy fluctuation in the right panel of figure \ref{fig:structure-entropy} displays cell-like convective patterns. These cell-like convective patterns are also observed in previously published literature \cite{kessar2019scale}. A wide range of average kinetic energy is covered in this comparison. The lower $Ra$ run has a kinetic energy of 27, and the higher $Ra$ run has a kinetic energy of 150. The large range in kinetic energy indicates that the inertial subrange for each simulation is different.
\begin{figure}[htb!]
		\centering
			\subfloat[]{\includegraphics[height=2.0in]{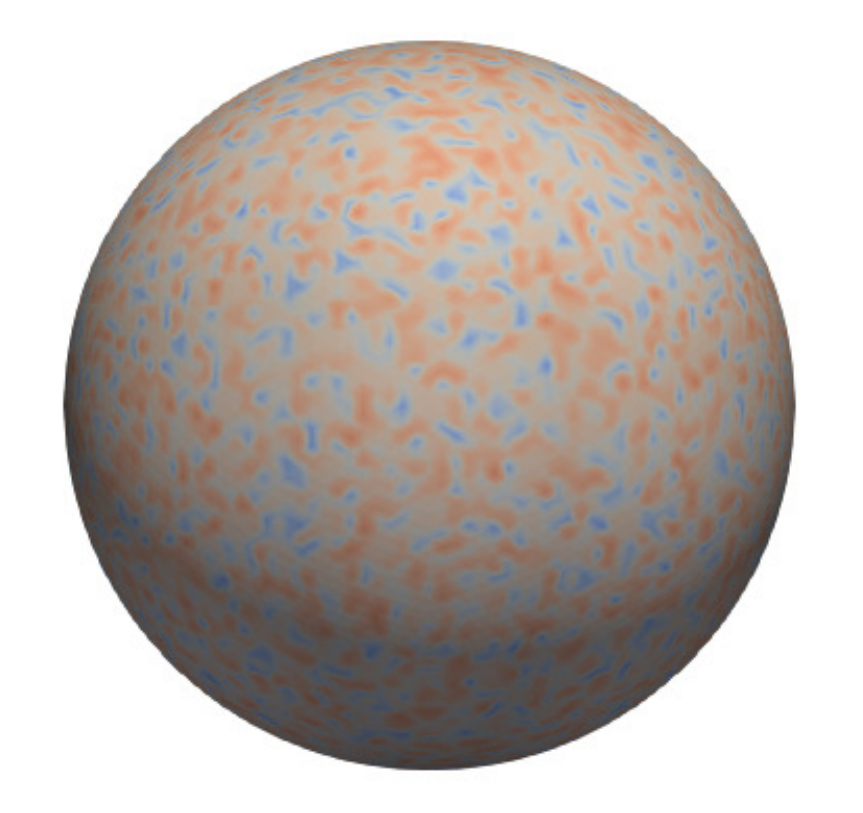}}
            \subfloat[]{\includegraphics[height=2.0in]{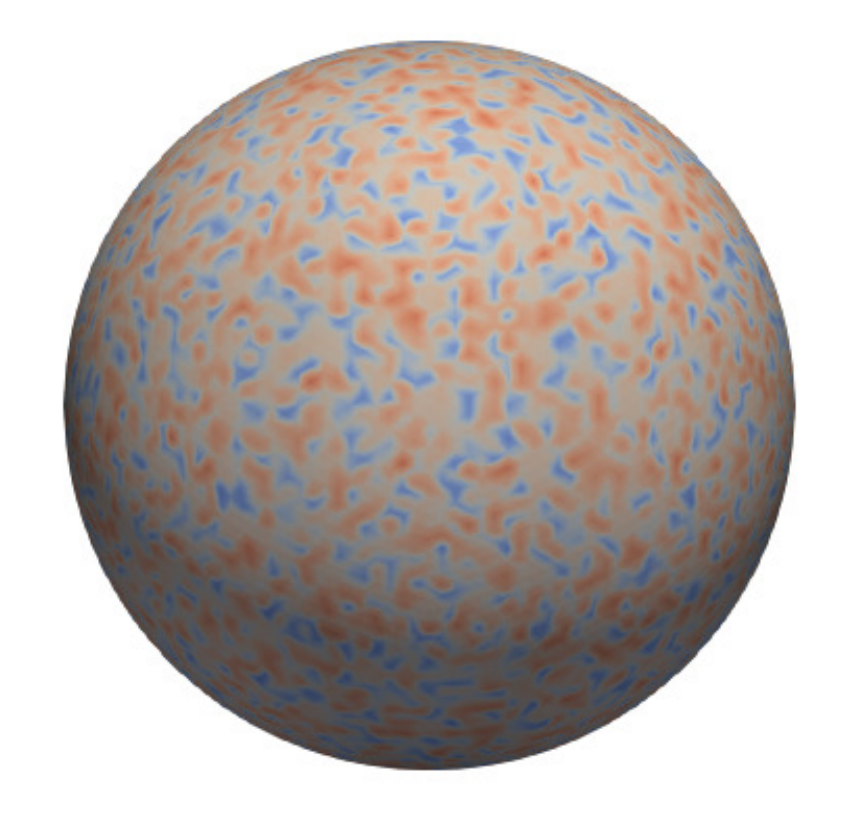}}
            \subfloat[]{\includegraphics[height=2.0in]{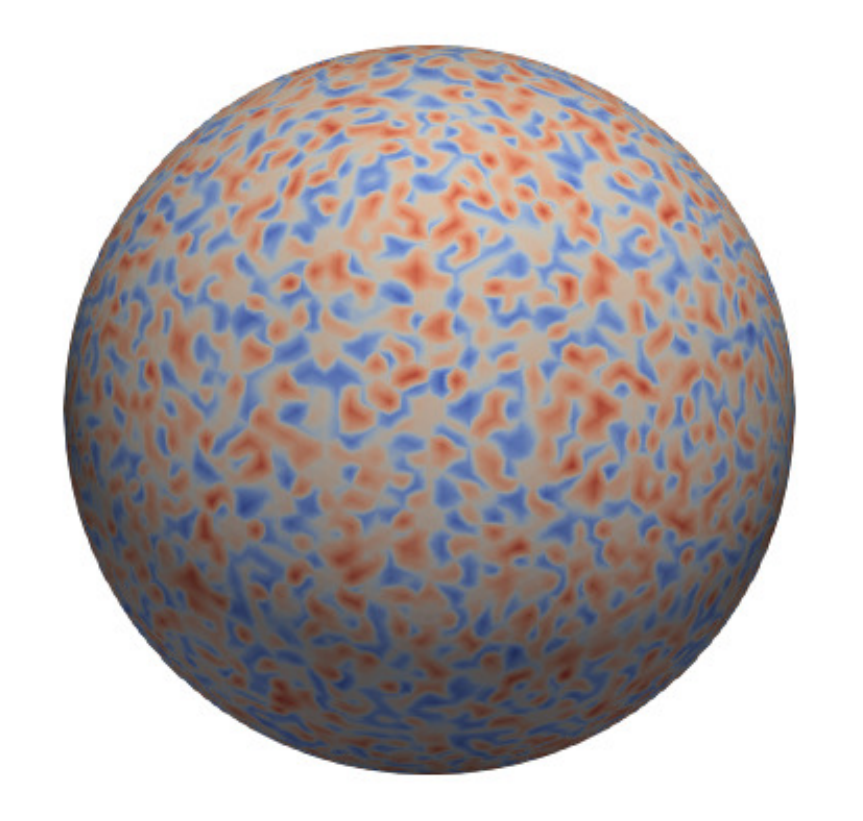}}
            \caption{Radial velocity at the outer layer of the spherical shells at $N_\rho$ = 2 for (a) $Ra$=3.6$\times$10$^3$, (b) $Ra$=5.6$\times$10$^3$, and (c) $Ra$=1.2$\times$10$^4$ (upflows in red and downflows in blue).} \label{fig:structure-ur}
\end{figure}
Convective patterns are also illustrated using radial velocity. Figure \ref{fig:structure-ur} shows the radial velocity at the outer layer of the spherical shells at $N_\rho = 2$ for $Ra$=3.6$\times$10$^3$, $Ra$=5.6$\times$10$^3$ and $Ra$=1.2$\times$10$^4$, where the red color indicates positive outflows and blue color indicate negative inflows. It is observed in Figure \ref{fig:structure-ur} that the flow structures appear to be a typical pattern of convection for a higher Rayleigh number as compared to a lower Rayleigh number. Convective patterns include broad upflows surrounded by narrow downflows. It can be seen that downflows at high $Ra$ are more intense and more connected than downflows at low $Ra$. With an increase in $Ra$ the general morphology of downflow pattern form a convective network of downflow lanes, which is slightly aligned with the cold plumes in Figure \ref{fig:structure-entropy}. A convective cell-like or granular pattern is more evident from entropy fluctuations in figure \ref{fig:structure-entropy}c. The observed behavior of convective flow patterns is in accordance with published literature \cite{guerrero2022implicit}.  

\subsection{Energy dynamics} \label{sec:energy}
\begin{figure}[htb!] 
		\centering
			\includegraphics[height=3.5in]{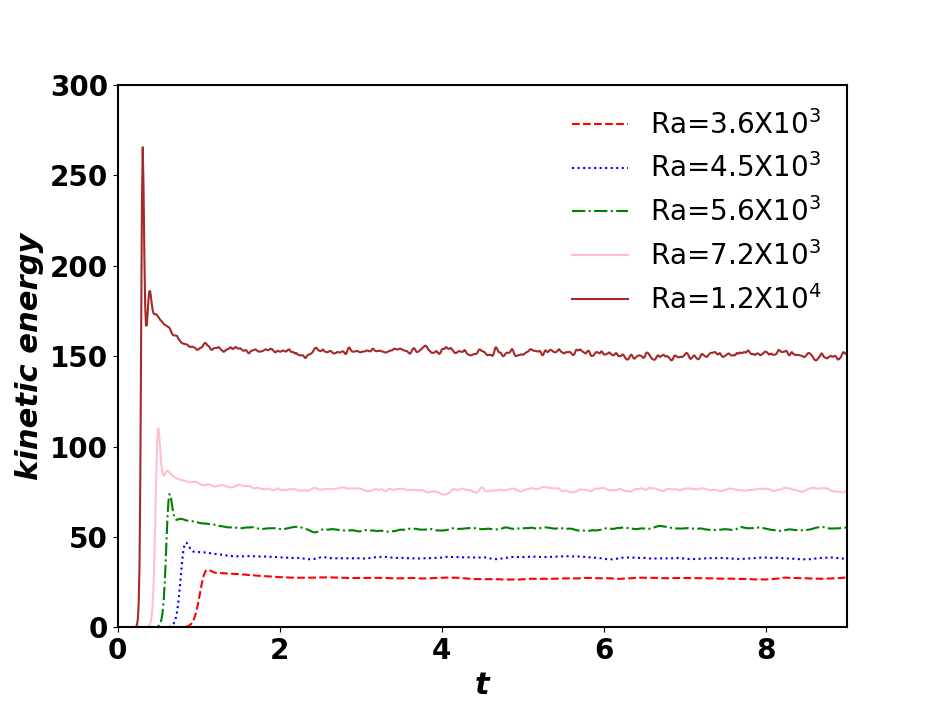}
            \caption{Time evolution of the average kinetic energy at $N_\rho$=2 for five different Rayleigh number.}\label{fig:ke-all}
\end{figure}

To strengthen our numerical investigation, we analyze convective flow quantitatively. Figure \ref{fig:ke-all} shows the volume average kinetic energy as a function of time at $N_\rho$=2 for five different $Ra$. 
\begin{figure}[htb!]
		\centering
			\includegraphics[height=3.5in]{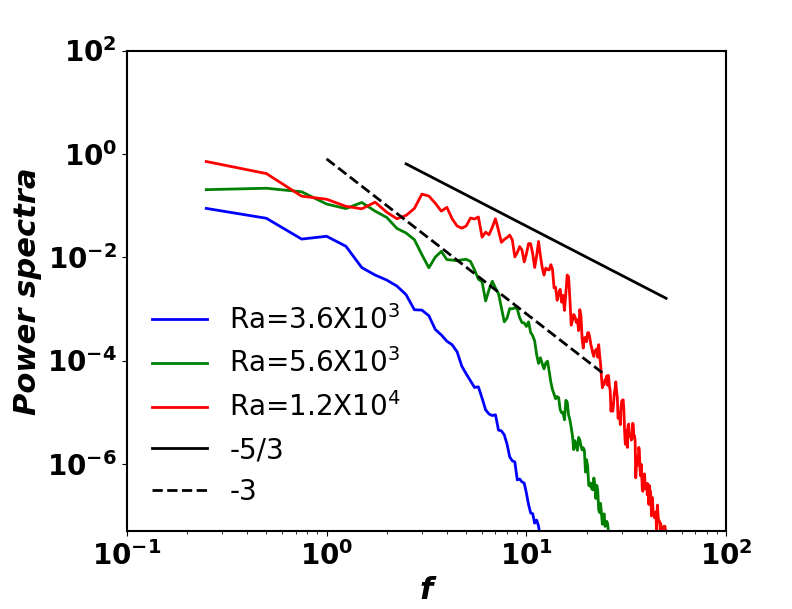}
            \caption{Convective power spectra computed from the average kinetic energy at $N_\rho$=2.}\label{fig:FFT-N2}
\end{figure}
It is observed that the initial development of convection instability occurs with a large spike in the interval $t$[0,2]. Later, the nonlinear saturation or statistical stationary state is achieved. With the increase in $Ra$, the convection instability occurs early with a larger spike and kinetic energy increases. We can see some fluctuation in the kinetic energy time series level in the saturated state at $Ra$=$1.2\times 10^4$; this conjectures the presence of chaotic convective motion. In comparison, the kinetic energy in the saturated state is stationary in time for lower $Ra$. The kinetic energy at $Ra$=$1.2\times 10^4$ is much larger and chaotic as compared to $Ra$=$3.6\times 10^3$, confirming our visualization of convective flow in Figure \ref{fig:structure-entropy}. Further, to check the convective motion, convective power spectra are calculated. Figure \ref{fig:FFT-N2} shows the convective power spectra of average kinetic energy. It is observed that at lower $Ra$ the power spectra follow -3 scaling. While at higher $Ra$, the power spectra show -5/3 scaling, indicating the convective flow is chaotic with granular patterns as seen in Figure \ref{fig:structure-entropy}. The inertial subrange -5/3 slope exhibits a small region, which elucidates that the convective flow is chaotic but not isotropic or homogeneous turbulent \cite{featherstone2016spectral,guerrero2022implicit}.

\subsection{Density ratio} \label{sec:density}
\begin{figure}[htb!]
		\centering
			\subfloat[]{\includegraphics[height=2.0in]{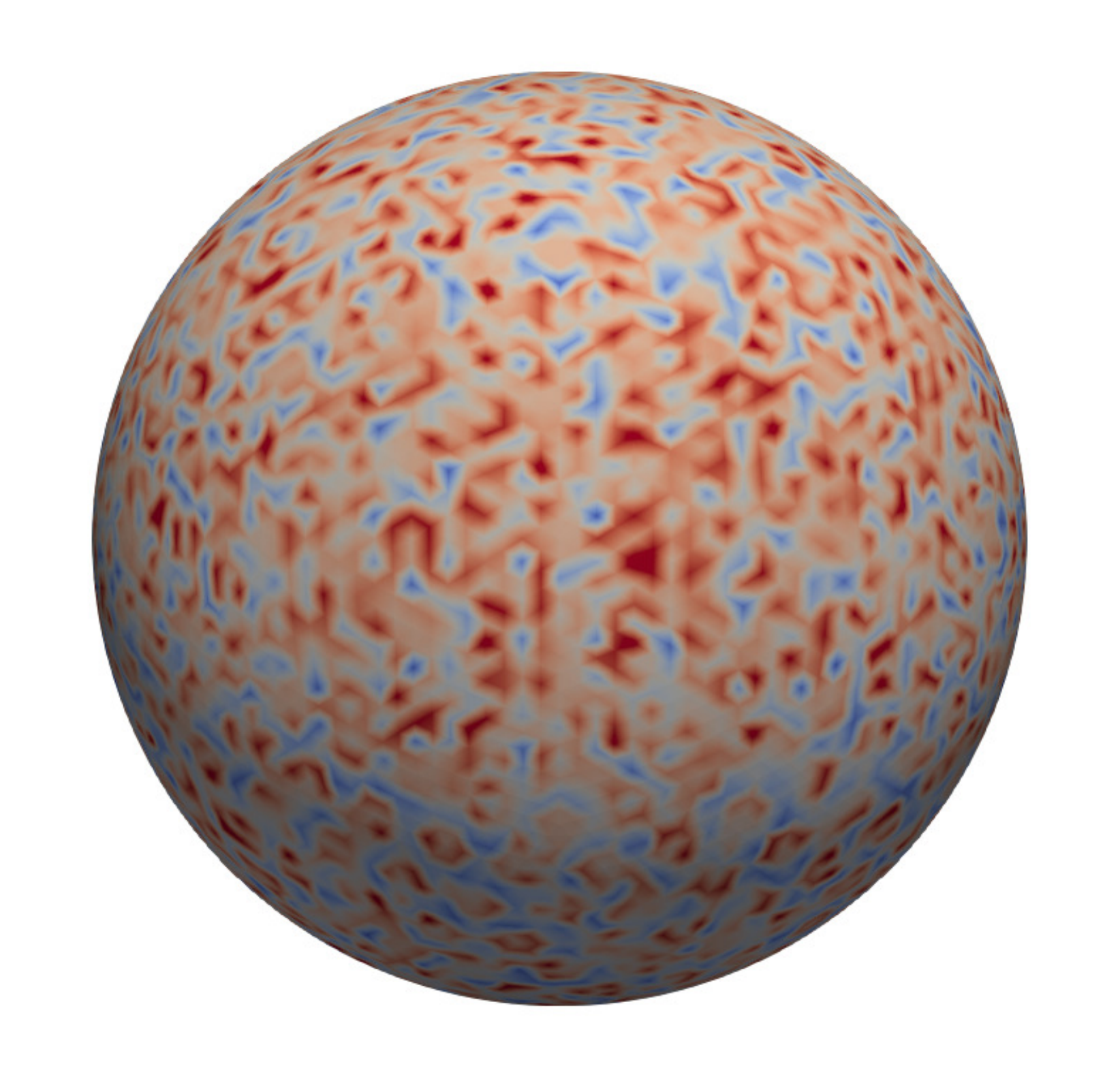}}
            \subfloat[]{\includegraphics[height=2.0in]{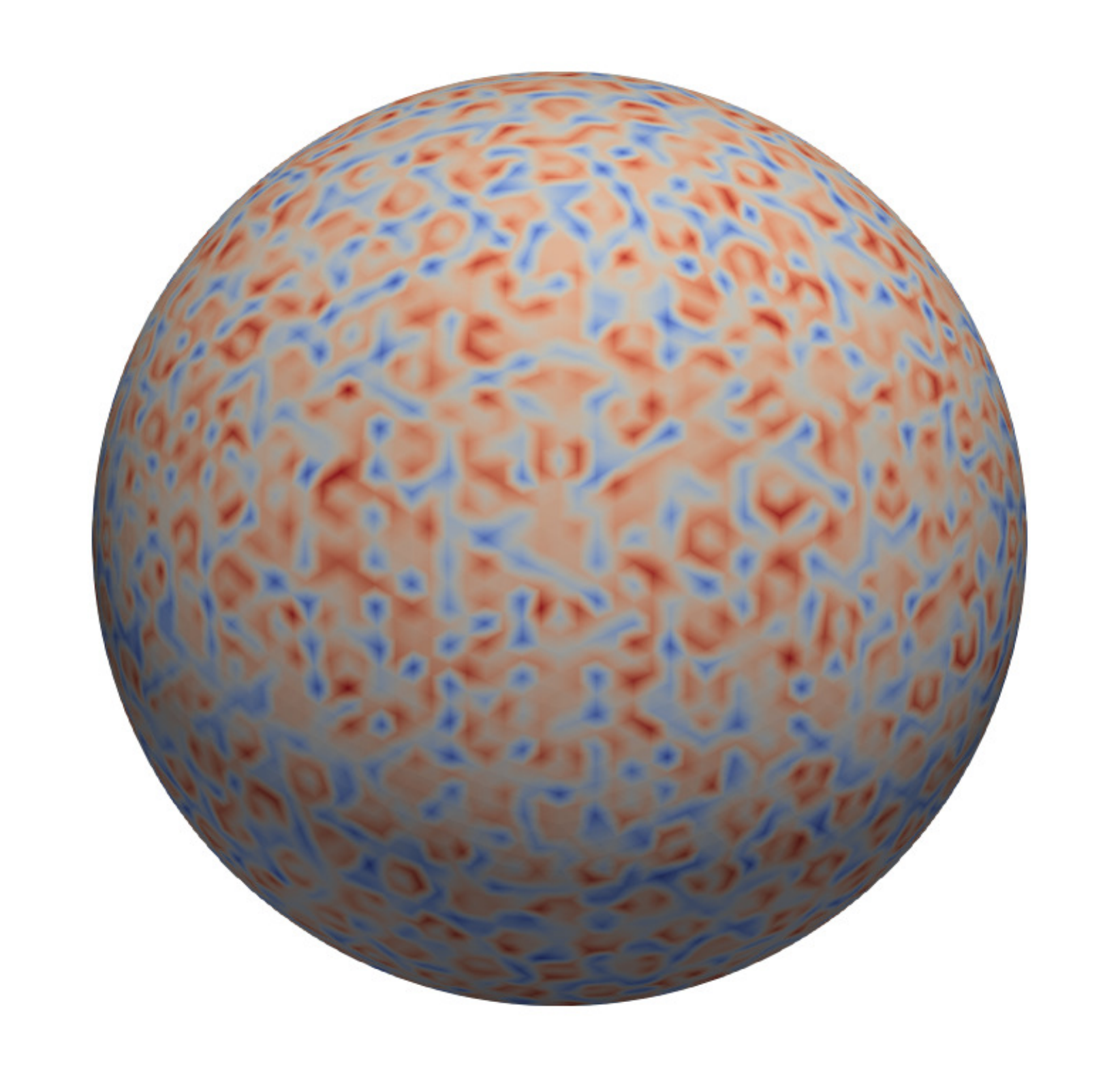}}
            \subfloat[]{\includegraphics[height=2.0in]{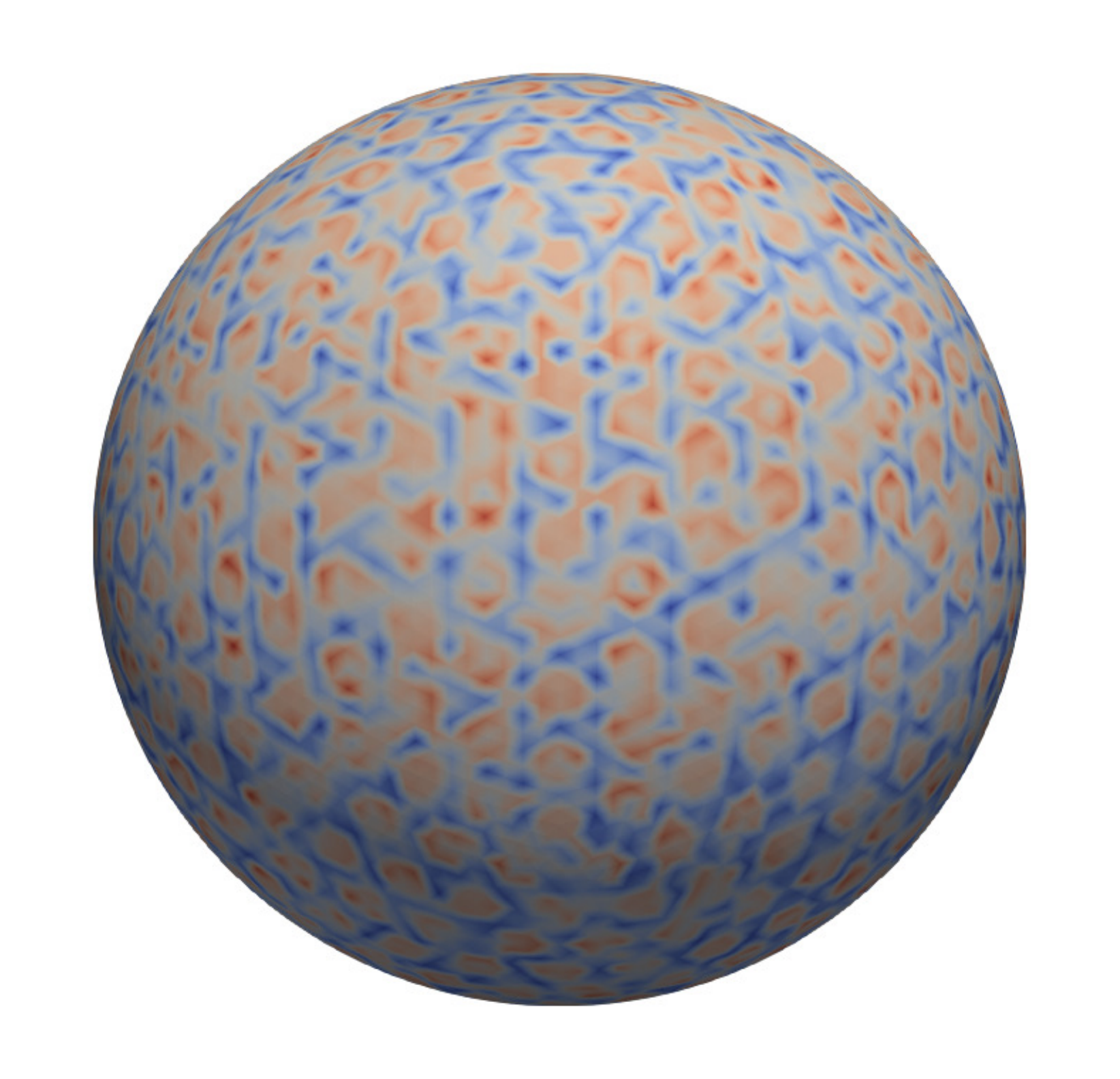}}
            \caption{Entropy fluctuations at the outer layer of the spherical shells at $Ra$=$1.2\times 10^4$ for (a) $N_\rho=0$, (b) $N_\rho=1$, and (c) $N_\rho=2$ (warm plumes in red and cold plumes in blue).} \label{fig:structure-density}
\end{figure}

In order to numerically test the effect of density ratio, we performed simulations at different density scale heights. Stratification influence on convective flow is demonstrated with entropy fluctuation on the spherical slice. Figure \ref{fig:structure-density} shows the entropy fluctuation at the outer layer of the spherical shells at $Ra$=$1.2\times 10^4$ for density scale height $N_\rho=0$-$2$, where warm upflows are in red and cool downflows are in blue.
\begin{figure}[htb!]
		\centering
			\includegraphics[height=3.5in]{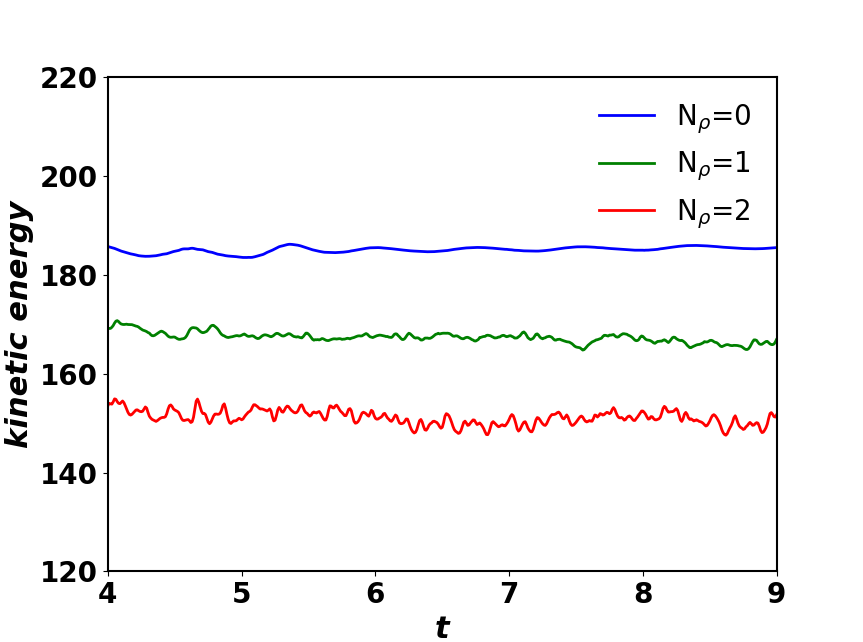}
            \caption{Average kinetic energy as a function of time for different density ratio at $Ra$=$1.2\times 10^4$.} \label{fig:ke-density}
\end{figure} 
\begin{figure}[htb!]
		\centering
            \includegraphics[height=3.6in]{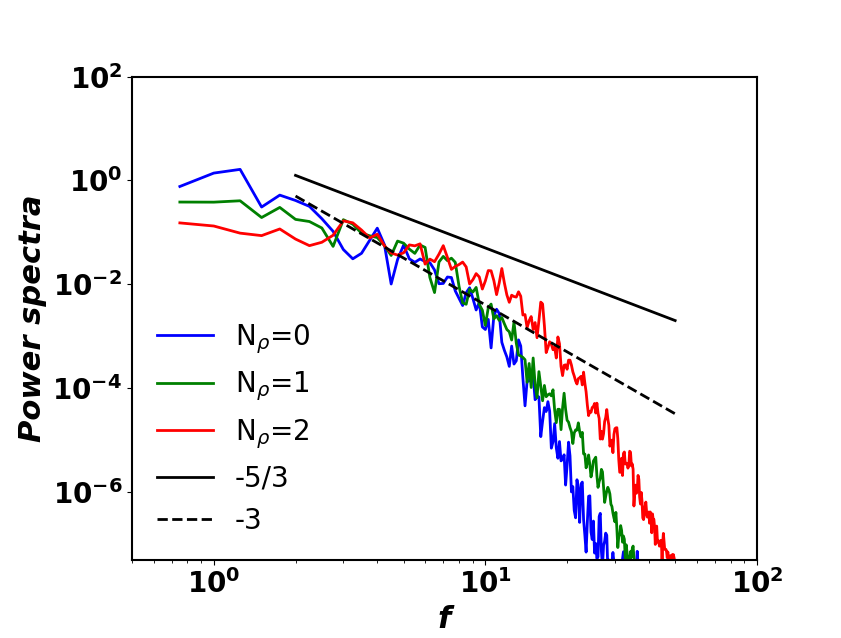}
            \caption{Convective power spectra computed from the average kinetic energy for different $N_\rho$ at $Ra$=$1.2\times 10^4$.} \label{fig:ke_density}
\end{figure}
For the density scale height $N_\rho=2$, a convective network or cell-like structure consisting of cold downflows is observed. As expected for $N_\rho=0$, it does not exhibit any cell-like structure. Indeed, there is a trend of cell-like structures or granular patterns with the increase in density ratio from $N_\rho=0$ to $2$. In qualitative terms, the present characteristics of structures follow the previously published work of solar convection \cite{bushby2012convectively,bushby2014mesogranulation,kessar2019scale}. However, the indication of granular patterns is further assessed through quantitative analysis of convective flow. Figure \ref{fig:ke-density} shows the average kinetic energy with time for density scale height $N_\rho=0-2$ at $Ra$ = 1.2 $\times 10^4$. It is observed that with an increase in density scale height, the kinetic energy decreases, while the fluctuation of kinetic energy increases in the saturated state. This behavior is consistent with previous studies of anelastic convection \cite{kessar2019scale}. To determine conclusively the coherent motion at a higher density ratio, we calculated the power spectra of kinetic energy. Figure \ref{fig:ke_density} shows spectra for $N_\rho$=0-2 at $Ra$=1.2 $\times 10^4$. There is a clear difference between the power spectra for $N_\rho$=0 and 2. At larger wavelengths, it is observed that with an increase in density scale height, the power spectra decreases. However, the energy distribution from large wavelength to small wavelength is much steeper for $N_\rho$=0 as compared to $N_\rho$=2. Therefore, the power spectra for higher density scale height follow -5/3 spectra, while the lower density scale height follows -3 spectra. For strong stratification, the spectra resemble an inertial range, which is attributed to chaotic convective motion. In other words, it indicates that the influence of cold plumes over cell-like structures increases, as displayed previously in Figure \ref{fig:structure-density}.  While, the shallower -5/3 slope does not resemble isotropic or homogeneous turbulence. Similar behavior of convective power spectra was earlier reported by Kessar et al. \cite{kessar2019scale} and Featherstone and Hindman \cite{featherstone2016spectral}. 

\section{Conclusions} \label{sec:conclusions}
The present work goal has been to develop, verify, and benchmark a hybrid discrete exterior calculus and finite difference (DEC-FD) solver for density-stratified thermal convection in spherical shells. The formulation and discretization of the anelastic equation using the hybrid DEC-FD method has been reported for the first time. We demonstrated the translation of anelastic partial differential equation (PDE) to smooth exterior calculus formulation and finally to DEC-FD discretized equation. We verified the discretized equation using the method of manufactured solutions (MMS), and it is observed that the discretized equation order of accuracy agrees with the theoretical order of accuracy. We performed numerical tests for anelastic thermal convection at density scale height $N_\rho$ =0 and $N_\rho >$0. The three-dimensional solar convection simulations are performed at moderate Rayleigh number, and present results agree with benchmark models. Subsequently, we demonstrated the spherical shells' convective cell-like or granular patterns and energy dynamics. In addition, we also examined the effect of density ratio on convective flow structures and energy dynamics. These investigations are an adequate test for a newly developed hybrid DEC-FD solver for density-stratified thermal convection because the results are satisfactory and agree with the published benchmark models.
\section*{Acknowledgments}
This publication is based upon work supported by the King Abdullah University of Science and Technology (KAUST) Office of Sponsored Research (OSR) under Award URF/1/4342-01. For computer time, this research used the Cray XC40, Shaheen II, of the Supercomputing Laboratory at King Abdullah University of Science $\&$ Technology (KAUST) in Thuwal, Saudi Arabia.
\section*{Appendix A}\label{appendixA}
The viscous heating term is discretized using the finite difference method on the spherical surface and radial direction. The viscous heating term $\Phi$ is defined as
\begin{equation}
\Phi=\frac{2\epsilon}{RaT_a\rho_a} \left[e_{ij} -\frac{1}{3}(\nabla\cdot\textbf{u})\delta_{ij}\right]^2
\end{equation}
where $e_{ij}$ is the strain rate tensor and it is defined as
\begin{equation}
    e_{ij}=\frac{1}{2}\left(\frac{\partial u_i}{\partial x_j}+\frac{\partial u_j}{\partial x_i}\right)
\end{equation}
The viscous heating term in cartesian coordinate is represented as
\begin{equation}\label{eq:cartesian}
\begin{split}
\Phi=\frac{2\epsilon}{RaT_a\rho_a} \bigg[\left(\frac{\partial u}{\partial x}\right)^2+\left(\frac{\partial v}{\partial y}\right)^2+\left(\frac{\partial w}{\partial z}\right)^2 
+ \frac{1}{2}\left(\frac{\partial u}{\partial y}+\frac{\partial v}{\partial x}\right)^2
+ \frac{1}{2}\left(\frac{\partial v}{\partial z}+\frac{\partial w}{\partial y}\right)^2 \\
+ \frac{1}{2}\left(\frac{\partial w}{\partial x}+\frac{\partial u}{\partial z}\right)^2 
- \frac{1}{3}\left(\frac{\partial u}{\partial x}+\frac{\partial v}{\partial y}+\frac{\partial w}{\partial z}\right)^2\bigg]
\end{split}
\end{equation}
The gradients of velocity components in equation (\ref{eq:cartesian}) are computed using a least squares fit.
\section*{Appendix B} \label{appendixB}
The residue term is obtained by solving the anelastic equations (\ref{eq0:continuity}) - (\ref{eq0:energy}) analytically using the manufactured solutions (see equations (\ref{eq:MS-ur})-(\ref{eq:MS-S})) . The residue term $R_c$, $R_m$ and $R_s$ for respective continuity, momentum and energy equation are
\begin{equation} 
R_c = \nabla \cdot({\rho}_{a} \bm{\widetilde{u}}) 
\end{equation}
\begin{equation}
R_m = \frac{\partial \bm{\widetilde{u}}}{\partial t} + (\bm{\widetilde{u}}\cdot\nabla)\bm{\widetilde{u}} + \nabla \frac{\tilde p}{\rho_a} - Ra Pr g(r) \tilde s e_r - \frac{Pr}{\rho_a}\left(\nabla ^2 \bm{\widetilde{u}} - \dfrac{1}{3} \nabla(\nabla\cdot \bm{\widetilde{u}})\right) 
\end{equation}
\begin{equation} 
R_s = \frac{\partial \tilde s}{\partial t} + (\bm{\widetilde{u}}\cdot\nabla)\tilde s - \chi(r)\tilde u_r - \frac{1}{\rho_a} \nabla^2 \tilde s - \frac{\epsilon}{RaT_a\rho_a} \left[e_{ij} -\frac{1}{3}(\nabla\cdot\bm{\widetilde{u}})\delta_{ij}\right]^2
\end{equation}
where $\bm{\widetilde{u}}$, $\tilde p$, and $\tilde s$ are the velocity vector, pressure, and entropy manufactured solution (MS), respectively. 
The MS velocity field vector $\bm{\widetilde{u}}$ is represented in spherical coordinates as $\tilde u_r$, $\tilde u_\theta$ and $\tilde u_\phi$. In the present work, we have the radial-momentum equation and surface-momentum equation. The radial-momentum equation residue term is represented as $R_{m,r}$. We recall that the surface velocity consists of $\tilde u_\theta$ and $\tilde u_\phi$. The manufactured solution $\tilde u_\theta$ is considered as a function of \textit{t}, \textit{r}, $\theta$ and $\phi$, while the manufactured solution $\tilde u_\phi$ is assumed constant. Subsequently, we considered that the residue term $R_{m,\perp}$ for the surface-momentum equation is calculated in $\theta$ direction while neglected in $\phi$ direction. The residue terms ($R_c$, $R_{m,r}$ , $R_{m,\perp}$ and $R_s$) in spherical coordinate are illustrated as follow
\begin{equation} \label{eq:R-c}
R_c = \frac{1}{r^2}\frac{\partial}{\partial r} \left(r^2\rho_a \tilde u_r\right) + \frac{\rho_a}{r \sin \theta} \frac{\partial \tilde u_\phi}{\partial \phi} + \frac{\rho_a}{r\sin \theta} \frac{\partial \sin \theta \tilde u_\theta}{\partial \theta} 
\end{equation}
\begin{equation} \label{eq:Rm-r}
\begin{split}
R_{m,r} = \frac{\partial \tilde u_r}{\partial t} + \tilde u_r\frac{\partial \tilde u_r}{\partial r} + \frac{\tilde u_\theta}{r}\frac{\partial \tilde u_r}{\partial \theta} + \frac{\tilde u_\phi}{r \sin \theta}\frac{\partial \tilde u_r}{\partial \phi} -\frac{\tilde u^2_\theta + \tilde u^2_\phi}{r} + \frac{1}{\rho_a}\frac{\partial \tilde p}{\partial r} - Ra Pr g(r) \tilde s e_r \\ 
- \frac{Pr}{\rho_a}\bigg[
\frac{\partial}{\partial r}\left(\frac{1}{r^2}\frac{\partial}{\partial r}(r^2 \tilde u_r)\right) + \frac{1}{r^2 \sin \theta}\frac{\partial}{\partial\theta}\left(\sin \theta \frac{\partial \tilde u_r}{\partial \theta}\right)
+\frac{1}{r^2 \sin^2 \theta}\frac{\partial^2 \tilde u_r}{\partial \phi^2} \\
-\frac{2}{r^2 \sin \theta}\frac{\partial}{\partial \theta} (\tilde u_\theta \sin \theta) - \frac{2}{r^2 \sin \theta}\frac{\partial \tilde u_\phi}{\partial \phi} - \dfrac{1}{3} \bigg(\frac{1}{r}\frac{\partial^2 \tilde u_r}{\partial \theta \partial r} + \frac{2}{r^2}\frac{\partial \tilde u_r}{\partial \theta} + \frac{1}{r^2}\frac{\partial^2 \tilde u_\theta}{\partial \theta^2}+\frac{1}{r^2}\frac{\partial^2 \tilde u_\theta}{\partial \theta^2}\\
+\frac{\cot \theta}{r^2}\frac{\partial \tilde u_\theta}{\partial \theta}-\frac{cosec^2 \theta}{r^2} \tilde u_\theta
-\frac{\cos \theta}{r^2 \sin^2 \theta} \frac{\partial \tilde u_\phi}{\partial \phi}+\frac{1}{r^2 \sin \theta}\frac{\partial^2 \tilde u_\phi}{\partial \theta \phi} + \frac{1}{r\sin \theta}\frac{\partial^2 \tilde u_r}{\partial \phi \partial r} \\
+ \frac{2}{r^2 \sin \theta}\frac{\partial \tilde u_r}{\partial \phi} + \frac{1}{r^2 \sin \theta}\frac{\partial^2 \tilde u_\theta}{\partial \theta \partial \phi} +\frac{\cos \theta}{r^2 \sin^2 \theta}\frac{\partial \tilde u_\theta}{\partial \phi}
+\frac{1}{r^2 \sin^2} \frac{\partial ^2 \tilde u_\phi}{\partial \phi^2}
\frac{\partial^2 \tilde u_r}{\partial^2 r} - \frac{2}{r^2} \tilde u_r\\
+\frac{2}{r}\frac{\partial \tilde u_r}{\partial r} - \frac{1}{r^2}\frac{\partial \tilde u_\theta}{\partial \theta}+\frac{1}{r^2}\frac{\partial^2 \tilde u_\theta}{\partial \theta \partial r}-\frac{\tilde u_\theta}{r^2}\cot \theta+\frac{\cot \theta}{r} \frac{\partial \tilde u_\theta}{\partial r}
-\frac{1}{r^2 \sin \theta} \frac{\partial \tilde u_\phi}{\partial \phi}+\frac{1}{r \sin \theta}\frac{\partial^2 \tilde u_\phi}{\partial r \partial \phi}\bigg)\bigg] 
\end{split}
\end{equation}
\begin{equation} \label{eq:Rm-theta}
\begin{split}
R_{m,\perp} = \frac{\partial \tilde u_\theta}{\partial t} + \tilde u_r\frac{\partial \tilde u_\theta}{\partial r} + \frac{\tilde u_\theta}{r}\frac{\partial \tilde u_\theta}{\partial \theta} + \frac{\tilde u_\phi}{r \sin \theta}\frac{\partial \tilde u_\theta}{\partial \phi} + \frac{\tilde u_r \tilde u_\theta}{r} -\frac{\tilde u^2_\phi \cot \theta}{r} + \frac{1}{r\rho_a}\frac{\partial \tilde p}{\partial \theta}\\
- \frac{Pr}{\rho_a}\bigg[
\frac{1}{r^2}\frac{\partial}{\partial r}\left(r^2\frac{\partial \tilde u_\theta}{\partial r}\right) + \frac{1}{r^2}\frac{\partial}{\partial\theta}\left(\frac{1}{\sin \theta} \frac{\partial }{\partial \theta}(\tilde u_\theta \sin \theta)\right)
+\frac{1}{r^2 \sin^2 \theta}\frac{\partial^2 \tilde u_\theta}{\partial \phi^2} -\frac{2}{r^2}\frac{\partial \tilde u_r}{\partial \theta} \\
- \frac{2\cot \theta}{r^2 \sin \theta}\frac{\partial \tilde u_\phi}{\partial \phi} - \dfrac{1}{3} \bigg(\frac{1}{r}\frac{\partial^2 \tilde u_r}{\partial \theta \partial r} + \frac{2}{r^2}\frac{\partial \tilde u_r}{\partial \theta} + \frac{1}{r^2}\frac{\partial^2 \tilde u_\theta}{\partial \theta^2}+\frac{1}{r^2}\frac{\partial^2 \tilde u_\theta}{\partial \theta^2}+\frac{\cot \theta}{r^2}\frac{\partial \tilde u_\theta}{\partial \theta}-\frac{cosec^2 \theta}{r^2} \tilde u_\theta\\
-\frac{\cos \theta}{r^2 \sin^2 \theta} \frac{\partial \tilde u_\phi}{\partial \phi}+\frac{1}{r^2 \sin \theta}\frac{\partial^2 \tilde u_\phi}{\partial \theta \phi} + \frac{1}{r\sin \theta}\frac{\partial^2 \tilde u_r}{\partial \phi \partial r} + \frac{2}{r^2 \sin \theta}\frac{\partial \tilde u_r}{\partial \phi} + \frac{1}{r^2 \sin \theta}\frac{\partial^2 \tilde u_\theta}{\partial \theta \partial \phi}\\
+\frac{\cos \theta}{r^2 \sin^2 \theta}\frac{\partial \tilde u_\theta}{\partial \phi} +\frac{1}{r^2 \sin^2} \frac{\partial ^2 \tilde u_\phi}{\partial \phi^2} +
\frac{\partial^2 \tilde u_r}{\partial^2 r} - \frac{2}{r^2} \tilde u_r+\frac{2}{r}\frac{\partial \tilde u_r}{\partial r} - \frac{1}{r^2}\frac{\partial \tilde u_\theta}{\partial \theta}+\frac{1}{r^2}\frac{\partial^2 \tilde u_\theta}{\partial \theta \partial r}\\
-\frac{\tilde u_\theta}{r^2}\cot \theta+\frac{\cot \theta}{r} \frac{\partial \tilde u_\theta}{\partial r} -\frac{1}{r^2 \sin \theta} \frac{\partial \tilde u_\phi}{\partial \phi}+\frac{1}{r \sin \theta}\frac{\partial^2 \tilde u_\phi}{\partial r \partial \phi}\bigg)\bigg] 
\end{split}
\end{equation}
\begin{equation} \label{eq:R-s}
\begin{split}
R_s = \frac{\partial \tilde s}{\partial t} + \tilde u_r\frac{\partial \tilde s}{\partial r} + \frac{\tilde u_\theta}{r}\frac{\partial \tilde s}{\partial \theta} + \frac{\tilde u_\phi}{r \sin \theta}\frac{\partial \tilde s}{\partial \phi} - \chi(r) \tilde u_r  - \frac{1}{\rho_a}\bigg[
\frac{1}{r^2}\frac{\partial}{\partial r}\left(r^2\frac{\partial \tilde s}{\partial r}\right) \\ 
+ \frac{1}{r^2 \sin \theta}\frac{\partial}{\partial\theta}\left(\sin \theta \frac{\partial \tilde s}{\partial \theta}\right)+\frac{1}{r^2 \sin^2 \theta}\frac{\partial^2 \tilde s}{\partial \phi^2} \bigg]  - \frac{2\epsilon}{RaT_a\rho_a} \bigg[\left(\frac{\partial \tilde u_r}{\partial r}\right)^2 + \left(\frac{1}{r}\frac{\partial \tilde u_\theta}{\partial \theta} + \frac{\tilde u_r}{r}\right)^2 \\ 
 + \left(\frac{1}{r \sin  \theta} \frac{\partial \tilde u_\phi}{\partial \phi} + \frac{\tilde u_r}{r} + \frac{\tilde u_\theta \cot \theta}{r}\right)^2
-\frac{1}{3}\bigg(\frac{\partial \tilde u_r}{\partial r} + \frac{\tilde u_r}{r} + \frac{1}{r} \frac{\partial \tilde u_\theta}{\partial \theta} + \frac{\tilde u_\theta \cot \theta}{r} \\
+ \frac{1}{r\sin \theta}\frac{\partial \tilde u_\phi}{\partial \phi}\bigg)^2\bigg] + \frac{\epsilon}{RaT_a\rho_a}\bigg[\left(\frac{\partial \tilde u_\theta}{\partial r} - \frac{\tilde u_\theta}{r} + \frac{1}{r} \frac{\partial \tilde u_r}{\partial \theta}\right)^2 
+ \bigg(\frac{1}{r}\frac{\partial \tilde u_\phi}{\partial \theta} - \frac{\tilde u_\phi \cot \theta}{r} \\ 
+ \frac{1}{r \sin \theta} \frac{\partial \tilde u_\theta}{\partial \phi}\bigg)^2 + \left( \frac{1}{r\sin\theta}\frac{\partial \tilde u_r}{\partial \phi} + \frac{\partial \tilde u_\phi}{\partial r} - \frac{\tilde u_\phi}{r}\right)^2\bigg]
\end{split}
\end{equation}
The manufactured solutions are substituted in the above equations (\ref{eq:R-c})-(\ref{eq:R-s}) and are solved analytically using the python (sympy \cite{negi2021train}) package.
\bibliographystyle{ieeetr}
\bibliography{article_paper}
\end{document}